\documentclass[a4paper,11pt]{article}
\pdfoutput=1
\usepackage{jheppub}
\usepackage[utf8]{inputenc}
\usepackage{epsfig,amsfonts,amsthm}
\usepackage{amsmath,amssymb}
\usepackage{float}
\usepackage{relsize}
\usepackage{xcolor}
\usepackage{tikz}

\newcommand{\be}{\begin{equation}}
\newcommand{\ee}{\end{equation}}
\newcommand{\bea}{\begin{eqnarray}}
\newcommand{\eea}{\end{eqnarray}}

\renewcommand{\Im}{\mathrm{Im }}

\newcommand{\triplet}[3]{ \left(\! \begin{array}{c}#1 \\ #2 \\ #3 \end{array}\!\right) }

\newcommand{\lr}[1]{ \langle #1 \rangle}

\newcommand{\Z}{\mathbb{Z}}
\newcommand{\mmatrix}[4]{ \left(\! \begin{array}{ccc}#1 & #2 \\ #3 & #4 \end{array}\!\right) }
\newcommand{\mmmatrix}[9]{ \left(\!\! \begin{array}{ccc}#1 & #2 & #3\\ #4 & #5 & #6\\ #7 & #8 & #9\\ \end{array}\!\!\right) }

\newcommand{\toCP}{\xrightarrow{CP}}

\def\lsim{\mathrel{\rlap{\lower4pt\hbox{\hskip1pt$\sim$}}
    \raise1pt\hbox{$<$}}}         %less than or approx. symbol
\def\gsim{\mathrel{\rlap{\lower4pt\hbox{\hskip1pt$\sim$}}
    \raise1pt\hbox{$>$}}}         %greater than or approx. symbol

\title{Limiting FCNC induced by a $CP$ symmetry of order 4}

\author[a,1]{Duanyang Zhao,}
\author[a,2]{Igor P. Ivanov,}
\author[b,3]{Roman Pasechnik,}
\author[a,4]{Pengming Zhang}
\affiliation[a]{School of Physics and Astronomy, Sun Yat-sen University, 519082 Zhuhai, China}
\affiliation[b]{Department of Physics, Lund University, SE-223 62 Lund, Sweden}
\emailAdd{zhaody8@mail2.sysu.edu.cn} 
\emailAdd{ivanov@mail.sysu.edu.cn} 
\emailAdd{roman.pasechnik@hep.lu.se} 
\emailAdd{zhangpm5@mail.sysu.edu.cn}

\abstract{
CP4 3HDM is a three-Higgs-doublet model based on the $CP$ symmetry of order 4 (CP4).
Imposing CP4 leads to remarkable connections between the scalar and Yukawa sectors
and unavoidably generates tree-level flavor-changing neutral couplings (FCNC).
It remains unclear whether FCNC can be sufficiently suppressed in the CP4 3HDM.
In this paper, we systematically explore this issue.
We first develop an efficient scanning procedure which takes the quark masses and mixing as input
and expresses the FCNC matrices in terms of physical quark observables
and quark rotation parameters.
This procedure allows us to explore the FCNC effects
for all the Yukawa sectors possible within the CP4 3HDM.
We find that, out of the eight possible CP4 Yukawa sectors,
only two scenarios are compatible with the $K$, $B$, $B_s$ and, 
in particular, $D$-meson oscillation constraints.
The results of this work serve as clear guidelines 
for future phenomenological scans of the model.
}

\begin{document}
\maketitle

\section{Introduction}

\subsection{Controlling FCNC in multi-Higgs-doublet models}

Multi-Higgs-doublet models, a popular framework for beyond the Standard Model (SM) constructions,
offer rich phenomenology from very few assumptions.
The two-Higgs-doublet model (2HDM) proposed by T.~D.~Lee half a century ago \cite{Lee:1973iz}
became a multi-Higgs golden standard
for collider searches \cite{Branco:2011iw,LHCHiggsCrossSectionWorkingGroup:2016ypw}.
The three-Higgs-doublet models (3HDMs), which offer more opportunities for model building than 2HDM, have also received much attention.
A version of 3HDM was first considered by S.~Weinberg in 1976 \cite{Weinberg:1976hu} as a means to combine natural flavor conservation,
and other works on various 3HDM-based fermion mass models quickly followed,
see \cite{Ivanov:2017dad} for a brief historical overview.
Today, the multi-Higgs-doublet framework remains an actively explored, phenomenologically attractive option.

Multi-Higgs models allow one to introduce global symmetries
acting on scalars and fermions. The symmetry groups and representation choices 
shape the scalar and Yukawa sectors and lead to remarkable,
structure-driven phenomenological consequences.
Within the 2HDM, a very popular choice is the softly broken $\Z_2$ symmetry \cite{Branco:2011iw},
which allows one to implement the natural flavor conservation principle~\cite{Glashow:1976nt,Paschos:1976ay,Peccei:1977hh}
and, as a result, to avoid the tree-level Higgs-mediated flavor changing neutral couplings (FCNC).

However, eliminating FCNCs altogether is not compulsory. Certain amount of tree-level FCNCs
can be tolerated if they are sufficiently suppressed and do not run into conflict with neutral meson oscillation parameters,
see \cite{Sher:2022aaa} for a recent review of various options.
For example, within the Cheng-Sher Ansatz \cite{Cheng:1987rs}, 
one assumes that the $(ij)$-entries of the quark-Higgs coupling matrices
are of the order of $\sqrt{m_i m_j}/v$. Although FCNCs are suppressed by the small quark masses,
this suppression seems to be insufficient \cite{Sher:2022aaa}.
Yet another approach is to look for models in which all flavor-violating transitions
are linked to a single source: the Cabibbo-Kobayashi-Maskawa (CKM) matrix.
This assumption known as the Minimal Flavor Violation hypothesis \cite{DAmbrosio:2002vsn} does not point to a unique model,
and various example have been constructed which realize it in some form.
In particular, the famous Branco-Grimus-Lavoura (BGL) model \cite{Branco:1996bq} offers a symmetry-based mechanism
in which FCNCs are controlled by products of small entries of CKM matrix.
Both the original BGL models and the 2HDMs which use generalizations 
of the BGL idea \cite{Botella:2014ska,Botella:2015hoa,Alves:2017xmk}
remain experimentally viable.

When attempting to control Higgs-induced FCNC effects on meson oscillation parameters,
one usually tries to keep individual quark-Higgs off-diagonal couplings suppressed.
However even if they are not as small as needed, one can invoke additional suppression mechanisms
which may exist in multi-Higgs models. 
Examples include partial compensation between $CP$-even and $CP$-odd Higgs boson exchanges \cite{Botella:2014ska}
and self-cancellation between the scalar and pseudoscalar couplings
of the same neutral Higgs boson \cite{Nebot:2015wsa}.
How efficient these additional mechanisms are depends, of course,
on the scalar sector and on the interplay between the off-diagonal couplings.

Within the scalar sector of the 2HDM, there is a very limited choice of global symmetry groups
\cite{Ivanov:2006yq,Nishi:2006tg,Ivanov:2007de,Maniatis:2007de,Ferreira:2009wh}.
When extended to the Yukawa sector \cite{Ferreira:2010ir,Ivanov:2013bka,Cogollo:2016dsd,Alves:2018kjr},
they constrain the Yukawa coupling matrices either insufficiently or too strongly \cite{Maniatis:2009vp,Ferreira:2010ir}.
3HDMs offer many more symmetry options for model building \cite{Ivanov:2012ry,Ivanov:2012fp,Darvishi:2019dbh},
including several non-abelian finite groups, as well as novel options for $CP$ symmetries and $CP$ violation.
In fact, the multi-Higgs activity in late 1970's and early 1980's was mainly driven by a trial-and-error search
for symmetry group and group representation assignments which would explain the emerging patterns
of the quark masses and mixing parameters.
Although no ideal candidate with an exact symmetry group was identified due to the reasons
eventually explained in \cite{Leurer:1992wg,Felipe:2013ie,Felipe:2014zka},
3HDMs with softly broken symmetries may do the job 
and remain today an actively explored direction \cite{Ivanov:2017dad,Bree:2023ojl}.

In the 3HDM, the magnitude of the quark-Higgs FCNC can be well controlled, too.
Already in the original proposal by Weinberg \cite{Weinberg:1976hu},
the tree-level FCNCs were absent due to the global $\Z_2\times \Z_2$ symmetry imposed.
Generic features of multi-Higgs-doublet models equipped with natural flavor conservation
were analyzed in \cite{Grossman:1994jb},
and many other works on specific 3HDMs models followed,
including the so-called democratic 3HDM \cite{Cree:2011uy}.
Recent examples of multi-Higgs-doublet models which are free from FCNC can be found in \cite{Yagyu:2016whx,deMedeirosVarzielas:2019dyu}.
Non-vanishing tree-level FCNC can also be tolerated if brought under control by additional symmetries,
for example, in the spirit of the BGL model \cite{Emmanuel-Costa:2017bti,Das:2021oik}
or via alignment of Yukawa matrices \cite{Penuelas:2017ikk}.

\subsection{CP4 3HDM: the present status}

Historically, multi-Higgs-doublet models emerged from the search for new opportunities for $CP$ violation,
\cite{Lee:1973iz,Weinberg:1976hu,Branco:1999fs}.
But they can also accommodate new forms of $CP$ symmetry,
which are physically distinguishable from the traditional $CP$.
It was noticed long ago that the action of discrete symmetries, such as $CP$,
on quantum fields is not uniquely defined \cite{feinberg-weinberg,Lee:1966ik,Branco:1999fs,weinberg-vol1}.
If a model contains several fields with identical gauge quantum numbers,
one can consider a general $CP$ transformation (GCP) which not only maps the fields to their conjugates
but also mixes them. For example, a GCP acting on complex scalar fields $\phi_i$, $i = 1, \dots, N$,
can be written as \cite{Ecker:1987qp,Grimus:1995zi}
\begin{equation}
\phi_i({\bf r}, t) \toCP {\cal CP}\,\phi_i({\bf r}, t)\, ({\cal CP})^{-1} = X_{ij}\phi_j^*(-{\bf r}, t), \quad X_{ij} \in U(N)\,.
\label{GCP}
\end{equation}
The conventional definition of $CP$ with $X_{ij} = \delta_{ij}$ is only one of many possible
choices and is, in fact, basis-dependent.
If a model does not respect the conventional $CP$ but is invariant under a GCP with a suitable matrix $X$,
then the model is explicitly $CP$-conserving \cite{Branco:1999fs}.

The presence of the matrix $X$ has consequences.
Applying the GCP twice leads to a family transformation, $\phi_i \mapsto (XX^*)_{ij}\phi_j$,
which may be non-trivial.
One may find that only when the GCP transformation is applied $k$ times that one arrives at the identity transformation;
thus, a $CP$ symmetry can be of order $k$.
The conventional $CP$ is of order 2; the next option is a $CP$ symmetry of order 4, denoted CP4.
Since the order of a transformation is basis-invariant, a model based on a CP4
represents a physically distinct $CP$-invariant model which cannot be achieved with the conventional $CP$.

Within the 2HDM, imposing CP4 on the scalar sector always leads to the usual $CP$ \cite{Ferreira:2009wh}.
In order to implement CP4 and avoid any conventional $CP$, one needs to pass to three Higgs doublets.
This model, dubbed CP4 3HDM, was constructed in \cite{Ivanov:2015mwl} building upon results of \cite{Ivanov:2011ae}
and was found to possess remarkable features which sometimes defy intuition \cite{Ivanov:2015mwl,Aranda:2016qmp,Haber:2018iwr}.
If CP4 remains unbroken at the minimum of the Higgs potential, it can protect the scalar dark matter candidates against decay
\cite{Ivanov:2018srm} and may be used to generate radiative neutrino masses \cite{Ivanov:2017bdx}.
Multi-Higgs-doublet models based on even higher order $CP$ symmetries were constructed in \cite{Ivanov:2018qni}.

CP4 symmetry can also be extended to the Yukawa sector leading to very particular patterns of the Yukawa matrices \cite{Ferreira:2017tvy}.
The CP4 transformation, by construction, mixes generations; therefore,
in order to avoid mass degenerate quarks, the explicit CP4 symmetry of the model must be spontaneously broken.
Then, the Yukawa sector contains enough free parameters to accommodate
the experimentally measured quark masses and mixing, as well as the appropriate amount of $CP$ violation.
The numerical scan of the scalar and Yukawa parameter spaces performed in \cite{Ferreira:2017tvy} yielded CP4 3HDM examples
which satisfied theoretical constraints, the electroweak precision tests, and did not violate
the kaon and $B$-meson oscillation properties.

Tree-level Higgs-mediated FCNCs are unavoidable in the CP4 3HDM.
The scalar alignment assumption used in \cite{Ferreira:2017tvy} guarantees that
the SM-like Higgs $h_{SM}$ does not change quark flavor.
But the other neutral scalars certainly do.
The scan performed in \cite{Ferreira:2017tvy} produced parameter space points
which satisfied the kaon and $B$-meson oscillation parameters.
However it remained unclear how exactly it happened: was it due to cancellation
among different Higgs contributions or was it a consequence of intrinsically
small FCNCs?
Also, $D$-meson oscillation constraints were not included in \cite{Ferreira:2017tvy}.

A few years later, Ref.~\cite{Ivanov:2021pnr} revealed that the charged Higgs boson couplings
to top quarks ruled out the vast majority of parameter space points believed viable in \cite{Ferreira:2017tvy}.
Almost all examples found in \cite{Ferreira:2017tvy}
contained one or two charged Higgs bosons lighter than the top quark,
which opened up non-standard top decay channels.
%This opens up non-standard top decay channels $t\to h_{1,2}^+ d_i$,
%where $d_i = (d, s, b)$, with the subsequent
%decays of the charged Higgses $h_{1,2}^+$ to lighter quarks.
Ref.~\cite{Ivanov:2021pnr} found that almost all these points conflict with the LHC Run 2 data
on searches for these top decays
%on searches for $BR(t \to h^+b)\times BR(h^+\to c\bar s)$ and $BR(t \to h^+b)\times BR(h^+\to c\bar b)$
or with the total top width.
Only a couple of points survived due to a very peculiar structure of their Yukawa matrices.
Once again, it remained unclear how these features emerged, what parameters controlled them,
and whether one could generate other benchmark models with custom-tailored charged Higgs coupling matrices.

Normally, one would want to repeat the parameter space scan including the new constraints.
However, an additional downside of \cite{Ferreira:2017tvy} was that only
a very small fraction of random scan points passed all the flavor constraints.
This was due to the intrinsically inefficient scanning method,
in which one begins with a random point, finds quark masses and mixing parameters way off
their experimental values, and then iteratively searches for a suitable combination of parameters
which would agree with quark masses and mixing values.

\subsection{The goals of the present work}

After Ref.~\cite{Ferreira:2017tvy}, CP4 3HDM emerged as a viable model
with remarkable cross-talk between the scalar and Yukawa sectors,
which accomplishes unexpectedly much for a 3HDM equipped with a single symmetry.
%and led to further predictions which needed to be explored.
At the same time, \cite{Ferreira:2017tvy} left many questions unanswered.
What is the typical magnitude of the FCNC in each class of Yukawa models?
How small the FCNCs can in principle become in each Yukawa sector, without compromising quark masses and mixing?
What parameters control their smallness?
Are there robust structural predictions for the amount of FCNC or for correlations among various entries?
Is there any efficient cancellation among different FCNC contributions to the meson oscillation parameters?

In order to answer all these questions, a new analysis of the CP4 3HDM phenomenology is definitely needed,
which must draw lessons from the above results.
This new analysis should be based on a more efficient scanning procedure in the Yukawa parameter space,
in which one takes the physical quark sector parameters as input and then, using the remaining freedom,
arrives at the FCNC matrices. We call this approach the inversion procedure.
Using it, we will be able to investigate the typical values of the off-diagonal FCNC entries
as well as the minimal values one could in principle achieve.

These are the main goals of the present work.
We will explicitly construct the inversion procedure for each type of the CP4 invariant Yukawa sector.
Using it, we will explore the magnitude of various FCNC couplings and check
whether they are compatible with the meson oscillation constraints
for neutral kaons, $B$, $B_s$, and $D$-meson systems.
To stay focused, we deal with quarks only, although leptons can be included in a similar fashion.

We stress that, in this work, we do not aim at a full parameter space scan of the model including the scalar sector.
Before such a scan can be undertaken, one first must develop a new procedure
and clarify the above issues on FCNC pattern.
The objective of this work is to understand the structural features of the Yukawa sector.
The results of this work, both the new procedure and the insights on the FCNCs,
will guide the full phenomenological scan of the CP4 3HDM,
which is delegated to a future publication.

The structure of the paper is the following.
In the next Section, we will remind the reader of the CP4 3HDM scalar and Yukawa sectors
and provide general expressions for the matrices $N_d$ and $N_u$
which describe coupling of neutral scalars with physical quarks.
Section~\ref{section-controlling} describes the inversion procedure, namely,
how one can recover the matrices $N_d$ and $N_u$
starting from the physical quark masses and mixing parameters.
In Section~\ref{section-FCNC-cases}, we apply this procedure for the three non-trivial
cases of the CP4 invariant Yukawa sectors.
Then, in Section~\ref{section-minimal}, we describe a toy model which helps
understand the minimal amount of FCNC effects and the parameters which control them.
Equipped with all these results, we describe in Section~\ref{section-numerical}
numerical results for FCNC in all CP4 3HDM Yukawa sectors.
In Section~\ref{section-discussion}, we discuss the grand picture which emerges
from this analysis and present guidelines on how to perform, in future,
a full parameter space scan which has a chance to satisfy all flavor constraints.
The Appendix contains technical details on the inversion procedure.

%%%%%%%%%%%%%%%%%%%%%%%%%%%%%%%%%%%%%%%%%%%%%%%%%%%%%%%%%%%%%%%%%%%%%%%%%%%%

\section{3HDM with $CP$-symmetry of order 4}\label{section-CP4}

\subsection{The scalar sector of CP4 3HDM}

The 3HDMs make use of three Higgs doublets $\phi_i$, $i = 1,2,3$ with identical quantum numbers.
CP4 acts on Higgs doublets by conjugation accompanied with a rotation in the doublet space.
Following \cite{Ivanov:2015mwl,Ferreira:2017tvy}, we use the following form of the CP4:
\begin{equation}
\phi_i \toCP X_{ij} \phi_j^*\,,\quad
X =  \left(\begin{array}{ccc}
1 & 0 & 0 \\
0 & 0 & i  \\
0 & -i & 0
\end{array}\right)\,.
\label{CP4-def}
\end{equation}
Applying this transformation twice leads to a non-trivial transformation in the space of doublets:
$\phi_{1} \mapsto \phi_{1}$, $\phi_{2,3} \mapsto -\phi_{2,3}$.
In order to get the identity transformation, we must apply CP4 four times, hence the order-4 transformation.
It is known that any CP-type transformation of order 4 acting in the space of three complex fields
can always be presented in the form \eqref{CP4-def} by a suitable basis change \cite{weinberg-vol1}.

The most general renormalizable 3HDM potential respecting this symmetry was presented in \cite{Ivanov:2015mwl}.
It can be written as
$V = V_0+V_1$, where
\begin{eqnarray}
V_0 &=& - m_{11}^2 (\phi_1^\dagger \phi_1) - m_{22}^2 (\phi_2^\dagger \phi_2 + \phi_3^\dagger \phi_3)
+ \lambda_1 (\phi_1^\dagger \phi_1)^2 + \lambda_2 \left[(\phi_2^\dagger \phi_2)^2 + (\phi_3^\dagger \phi_3)^2\right]
\nonumber\\
&+& \lambda_3 (\phi_1^\dagger \phi_1) (\phi_2^\dagger \phi_2 + \phi_3^\dagger \phi_3)
+ \lambda'_3 (\phi_2^\dagger \phi_2) (\phi_3^\dagger \phi_3)\nonumber\\
&+& \lambda_4 \left[(\phi_1^\dagger \phi_2)(\phi_2^\dagger \phi_1) + (\phi_1^\dagger \phi_3)(\phi_3^\dagger \phi_1)\right]
+ \lambda'_4 (\phi_2^\dagger \phi_3)(\phi_3^\dagger \phi_2)\,,
\label{V0}
\end{eqnarray}
with all parameters real, and
\begin{equation}
V_1 = \lambda_5 (\phi_3^\dagger\phi_1)(\phi_2^\dagger\phi_1) +
\lambda_8(\phi_2^\dagger \phi_3)^2 + \lambda_9(\phi_2^\dagger\phi_3)(\phi_2^\dagger\phi_2-\phi_3^\dagger\phi_3) + h.c.
\label{V1}
\end{equation}
with real $\lambda_5$ and complex $\lambda_8$, $\lambda_9$.
Out of the two remaining complex parameters, only one can be made real by a suitable rephasing of $\phi_2$ and $\phi_3$.
However, in this work we prefer to keep $\lambda_8$, $\lambda_9$ complex
because we will rely on the residual rephasing freedom
to make the vacuum expectation values (vevs) real.
Also, two additional terms with coefficients $\lambda_6$ and $\lambda_7$ invariant under the same CP4
could be inserted in \eqref{V1} but, without loss of generality,
they can be eliminated with a suitable CP4 preserving rotation, see details in \cite{Ferreira:2017tvy}.

Minimization of this potential and the resulting scalar bosons mass matrices were studied in \cite{Ferreira:2017tvy}.
In order to avoid pathologies in the quark sector, CP4 must be spontaneously broken.
The three doublets acquire vevs, which can in principle be complex,
but the rephasing freedom allows us to render the three vevs real:
\begin{equation}
\lr{\phi_i^0} = \frac{1}{\sqrt{2}}(v_1,\, v_2,\, v_3)
\equiv \frac{v}{\sqrt{2}}\, (c_\beta, s_\beta c_\psi, s_\beta s_\psi)\,.\label{vevs}
\end{equation}
With $v = 246$ GeV fixed, the position of the minimum is described by two angles $\beta$ and $\psi$
(in the expression above, we used the shorthand notation for sines and cosines of these angles).
Spontaneous breaking of CP4 leads to four physically equivalent minima
linked by the CP4 transformation. Since we are allowed to pick up any of them,
we can safely assume that the angle $\psi$ lies in the first quadrant.

In the standard minimization procedure, one begins with the scalar potential $V$ and locates the minimum.
For a phenomenological analysis, it is more convenient to choose angles $\beta$ and $\psi$ are free parameters.
In this way, there will be two relations among the parameters of the potential, which could also be found in \cite{Ferreira:2017tvy}.
For example, knowing $\psi$ allows us to relate the imaginary parts of $\lambda_8$ and $\lambda_9$:
$s_{2\psi} \Im(\lambda_8) + c_{2\psi} \Im(\lambda_9) = 0$.

Next, it is convenient to rotate the three doublets to a Higgs basis as
\begin{equation}
\triplet{H_1}{H_2}{H_3} =
\mmmatrix{c_\beta}{s_\beta c_\psi}{s_\beta s_\psi}{-s_\beta}{c_\beta c_\psi}{c_\beta s_\psi}{0}{-s_\psi}{c_\psi}
\triplet{\phi_1}{\phi_2}{\phi_3},
\quad
\triplet{\phi_1}{\phi_2}{\phi_3} =
\mmmatrix{c_\beta}{-s_\beta}{0}{s_\beta c_\psi}{c_\beta c_\psi}{-s_\psi}{s_\beta s_\psi}{c_\beta s_\psi}{c_\psi}
\triplet{H_1}{H_2}{H_3} \,.
\label{matrix-P}
\end{equation}
In this Higgs basis, the vev is located only in $H_1$:
\begin{equation}
\lr{H_1^0} = \frac{v}{\sqrt{2}}\,, \quad \lr{H_2^0} = \lr{H_3^0} = 0\,,
\end{equation}
The would-be Goldstone modes populate $H_1$, while all fields in the doublets $H_2$, $H_3$ are physical scalars degrees of freedom.
Expanding the potential near the minimum, we obtain five neutral scalar bosons and two pairs of charged Higgses $h^{\pm}_{1,2}$.
In general, all neutral Higgs bosons can couple to $WW$ and $ZZ$ pairs.
However, if one fixes $m_{11}^2 = m_{22}^2$, the model exhibit scalar alignment: one of the neutral Higgses
$h_{SM}$ couples to the $WW$ and $ZZ$ exactly as in the SM,
while the other four neutral bosons $h_2$ through $h_5$ decouple from these channels.

Scalar alignment is not a necessary assumption. Run 2 LHC results on the 125 GeV Higgs boson properties
leave room for $\sim 10\%$ deviations in the $hWW/hZZ$ and $hf\bar f$ couplings \cite{CMS:2022dwd,ATLAS:2022vkf}.
However if unsuppressed FCNC couplings leak into the SM-like Higgs interactions, they may immediately run in conflict 
with the meson oscillation
constraints and with the Higgs-top FCNC measurements.
What amount of scalar misalignment is still tolerable is a compex issue as it depends on the interplay between the FCNC matrices,
the exact neutral scalars mixing matrix, and the SM-like Higgs interactions.
We definitely plan to investigate these issues in future. What we do in the present work ---
exploring generic FCNC couplings allowed within the CP4 3HDM --- can be seen
as the starting point to this study.
Thus, for now, we adopt scalar alignment, and this is all the information from the scalar sector we need in this work.

\subsection{General FCNC matrices in 3HDM}

To set up the notation, let us begin with the general expressions for the quark Yukawa sector in the 3HDM:
\begin{equation}
-{\cal L}_Y = \bar{Q}^0_L (\Gamma_1 \phi_1 + \Gamma_2 \phi_2 + \Gamma_3 \phi_3) d_R^0 +
\bar{Q}^0_L (\Delta_1 \tilde\phi_1 + \Delta_2 \tilde\phi_2 + \Delta_3 \tilde\phi_3) u_R^0 + h.c.\label{Yukawa-general}
\end{equation}
Here, we use the notation of \cite{Botella:2018gzy} extended to the 3HDM.
The three generations of quarks are implicitly assumed everywhere, their indices suppressed for brevity.
The superscript $0$ for the quark fields indicates that these are the starting quark fields;
when we pass to the physical quark fields by diagonalizing the quark mass matrices, we will remove this superscript.
Using the real vev basis \eqref{vevs}, we write the quark mass matrices as
\begin{equation}
M_d^0 = \frac{v}{\sqrt{2}}(\Gamma_1 c_\beta + \Gamma_2 s_\beta c_\psi + \Gamma_3s_\beta s_\psi)\,,
\quad
M_u^0 = \frac{v}{\sqrt{2}}(\Delta_1 c_\beta + \Delta_2 s_\beta c_\psi + \Delta_3s_\beta s_\psi)\,.
\label{Md0Mu0-general}
\end{equation}
They are, in general, non-diagonal and complex.
The interaction of the neutral (complex) scalars with the quarks can be described
both in the initial basis and in the Higgs basis we chose; the relation between the two bases is given by
\begin{equation}
\Gamma_1 \phi_1^0 + \Gamma_2 \phi_2^0 + \Gamma_3 \phi_3^0 =
\frac{\sqrt{2}}{v} (H_1^0 M_d^0 + H_2^0 N_{d2}^0 + H_3^0 N_{d3}^0)\,,
\end{equation}
where
\begin{eqnarray}
N_{d2}^0 &=& M_d^0 \cot\beta -\frac{v}{\sqrt{2} s_\beta} \Gamma_1
=  - M_d^0 \tan\beta + \frac{v}{\sqrt{2}c_\beta}(\Gamma_2 c_\psi + \Gamma_3 s_\psi)\,,\nonumber\\
N_{d3}^0 &=& \frac{v}{\sqrt{2}}(-\Gamma_2 s_\psi + \Gamma_3 c_\psi)\,.\label{Nd20Nd30-general}
\end{eqnarray}
For the up-quark sector, we obtain
\begin{equation}
\Delta_1 (\phi_1^0)^* + \Gamma_2 (\phi_2^0)^* + \Delta_3 (\phi_3^0)^* =
\frac{\sqrt{2}}{v} [(H_1^0)^* M_u^0 + (H_2^0)^* N_{u2}^0 + (H_3^0)^* N_{u3}^0]\,,
\end{equation}
where
\begin{eqnarray}
N_{u2}^0 &=& M_u^0 \cot\beta - \frac{v}{\sqrt{2} s_\beta} \Delta_1
=  - M_u^0 \tan\beta + \frac{v}{\sqrt{2}c_\beta}(\Delta_2 c_\psi + \Delta_3 s_\psi)\,,\nonumber\\
N_{u3}^0 &=& \frac{v}{\sqrt{2}}(-\Delta_2 s_\psi + \Delta_3 c_\psi)\,.
\end{eqnarray}
As usual, these mass matrices are diagonalized by unitary transformations of the quark fields,
\begin{equation}
d_L^0 = V_{dL} d_L\,, \quad d_R^0 = V_{dR} d_R\,, \quad
u_L^0 = V_{uL} u_L\,, \quad u_R^0 = V_{uR} u_R\,.\label{quark-rotations}
\end{equation}
which lead to the CKM matrix $V_{\rm CKM} = V_{uL}^\dagger V_{dL}$ and
\begin{equation}
D_d = V_{dL}^\dagger M_d^0 V_{dR} = {\rm diag}(m_d, m_s, m_b)\,,\quad
D_u = V_{uL}^\dagger M_u^0 V_{uR} = {\rm diag}(m_u, m_c, m_t)\,.
\end{equation}
The same quark rotation matrices also act on the matrices $N$:
\begin{equation}
N_{d2} = V_{dL}^\dagger N_{d2}^0 V_{dR}\,, \quad N_{u2} = V_{uL}^\dagger N_{u2}^0 V_{uR}
\end{equation}
and, similarly, for $N_{d3}$, $N_{u3}$.

The four Yukawa matrices $N_{d2}$, $N_{d3}$, $N_{u2}$, $N_{u3}$ are the key objects we study in this work.
They describe the coupling patterns
of the neutral complex fields $H_2^0$ and $H_3^0$ with the three generations of physical quarks.
Their off-diagonal elements indicate the strength of FCNC.
Within the 2HDM, we would only get $N_{d2}$ and $N_{u2}$, see e.g. \cite{Botella:2018gzy}.
The additional matrices $N_{d3}$ and $N_{u3}$ arise in the 3HDM,
and their patterns can be very different from $N_{d2}$ and $N_{u2}$.

Notice that the complex fields $H_2^0$ and $H_3^0$ contain four real neutral degrees of freedom.
All these components mix, and to get the physical Higgses, we would need to diagonalize
the neutral scalar mass matrix.
In the present work, we do not aim at a full phenomenological scan of the model;
we only want understand how FCNCs can in principle be limited within the CP4 3HDM.
For that purpose, it is sufficient and more transparent to work directly with
$N_{d2}$, $N_{d3}$ and $N_{u2}$, $N_{u3}$ without invoking mixing among neutral bosons.

\subsection{The Yukawa sector of the CP4 3HDM}

CP4 symmetry can be extended to the Yukawa sector of 3HDM.
This extension is not unique, but there is a limited number of structurally different options.
This problem was solved in \cite{Ferreira:2017tvy} and yielded four distinct cases, labeled $A$, $B_1$, $B_2$, and $B_3$,
separately in the up and down-quark sectors.
For example, in the down-quark sector, case $A$ represents the trivial solution
	\begin{equation}
\Gamma_1 = \mmmatrix{g_{11}}{g_{12}}{g_{13}}%
{g_{12}^*}{g_{11}^*}{g_{13}^*}%
{g_{31}}{g_{31}^*}{g_{33}}\,,\quad
\Gamma_{2,3} = 0\,,\label{caseA}
\end{equation}
which is completely free from FCNCs.
Cases $B_1$, $B_2$, $B_3$ involve all three Yukawa matrices with a non-trivial structure:
\begin{itemize}
	\item
	Case $B_1$:
	\begin{equation}
	\Gamma_1 = \mmmatrix{0}{0}{0}{0}{0}{0}{g_{31}}{g_{31}^*}{g_{33}}\,,\quad
	\Gamma_2 = \mmmatrix{g_{11}}{g_{12}}{g_{13}}{g_{21}}{g_{22}}{g_{23}}{0}{0}{0}\,,\quad
	\Gamma_3 =  \mmmatrix{-g_{22}^*}{-g_{21}^*}{-g_{23}^*}{g_{12}^*}{g_{11}^*}{g_{13}^*}{0}{0}{0}\,.
	\label{caseB1}
	\end{equation}
	
	\item
	Case $B_2$:
	\begin{equation}
	\Gamma_1 = \mmmatrix{0}{0}{g_{13}}{0}{0}{g_{13}^*}{0}{0}{g_{33}}\,,\quad
	\Gamma_2 = \mmmatrix{g_{11}}{g_{12}}{0}{g_{21}}{g_{22}}{0}{g_{31}}{g_{32}}{0}\,,\quad
	\Gamma_3 =  \mmmatrix{g_{22}^*}{-g_{21}^*}{0}{g_{12}^*}{-g_{11}^*}{0}{g_{32}^*}{-g_{31}^*}{0}\,.
	\label{caseB2}
	\end{equation}
	
	\item
	Case $B_3$:
	\begin{equation}
	\Gamma_1 = \mmmatrix{g_{11}}{g_{12}}{0}{-g_{12}^* }{g_{11}^*}{0}{0}{0}{g_{33}}\,,\quad
	\Gamma_2 = \mmmatrix{0}{0}{g_{13}}{0}{0}{g_{23}}{g_{31}}{g_{32}}{0}\,,\quad
	\Gamma_3 = \mmmatrix{0}{0}{-g_{23}^*}{0}{0}{g_{13}^*}{g_{32}^*}{-g_{31}^*}{0}\,.
	\label{caseB3}
	\end{equation}
\end{itemize}
In all cases, all the parameters apart from $g_{33}$ can be complex.
A similar set of cases is found in the up-quark sector.

Since the up and down sectors involve the same left-handed doublets,
one can only combine cases $A$ or $B_2$ in the up sector with $A$ or $B_2$ in the down sectors,
and cases $B_1$ or $B_3$ in the up sector with $B_1$ or $B_3$ in the down sector.
This leads to eight possible pairings.
However one of them, $(A,A)$, should be disregarded:
as shown in the Appendix, it is unable to generate the $CP$-violating phase in the CKM matrix.
Thus, we are left with seven viable combinations for the CP4 invariant Yukawa sectors:
%which can be viewed as CP4 3HDM counterparts of the Type-I/Type-II classification:
\begin{eqnarray}
\mbox{(down, up):} && (B_1, B_1)\,, \quad (B_1, B_3)\,, \quad (B_3, B_1)\,, \quad (B_3, B_3)\,, \label{cases-first-group}\\
&& (A, B_2)\,, \quad (B_2, A)\,, \quad (B_2, B_2)\,. \label{cases-second-group}
\end{eqnarray}
Since case $(A,A)$ is excluded, we arrive at the important conclusion that FCNCs are unavoidable in CP4 3HDM.

%%%%%%%%%%%

\section{Controlling FCNC in the CP4 3HDM: the strategy}\label{section-controlling}

\subsection{The inversion procedure}\label{subsection-inversion}

\begin{figure}[!h]
	\begin{center}
	\begin{tikzpicture}
	\large
	\node (GD) at (-0.5,1) {$\Gamma_i$, $\Delta_i$};
	\node (M0) at (2,1) {$M_d^0$, $M_u^0$};
	\node (M) at (5,1) {$m_q$, $V_{\rm CKM}$};
	\node (N0) at (2,-0.5) {$N_d^0$, $N_u^0$};
	\node (N) at (5,-0.5) {$N_d$, $N_u$};
	\draw[->] (GD) -- node[below] {$v_i$} (M0); \draw[->] (M0) -- node[below] {\scriptsize diag} (M);
	%\draw[->] (GD) |- (N0);
	\draw[->] (GD) -- (1.3, -0.3);
	\draw[->] (N0) -- (N);
	\draw[white!60!blue,line width=1mm,->] (2,1.5) arc [start angle=20, end angle=160, x radius=1.4, y radius=.3];
	\draw[white!60!blue,line width=1mm,->] (5,0.8) -- (5,-0.3);
	\node[black!50!blue] at (0.7,2.1) {inversion};
	
	%\node at (0.9,0.5) {$v_i$};
	\node[blue] at (6.9,0.3) {$V_{dL}, V_{dR}, V_{uL}, V_{uR}$};
%	\node[blue] at (5.1,0.2) {\scriptsize $V_{uL}, V_{uR}$};
	\end{tikzpicture}
	\caption{\label{Fig:diagram} In a generic multi-Higgs model, one begins with $\Gamma_i$, $\Delta_i$,
		computes the mass matrices and diagonalizes them. This procedure, indicated by thin arrows, is usually irreversible.
		In certain models, one can perform inversion (thick light blue arrows),
		which allows one to pass directly from the quark properties to the FCNC matrices.}
	\end{center}
\end{figure}
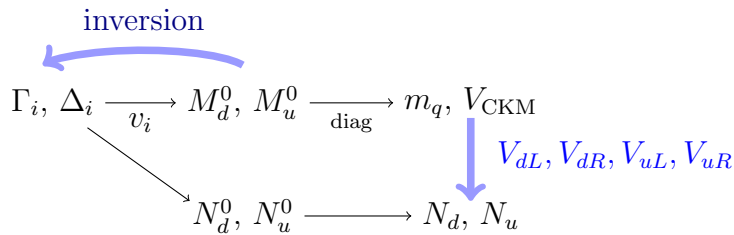

Fitting the quark masses and the parameters of the CKM mixing matrix is, in general,
a non-trivial task for multi-Higgs-doublet models with generic Yukawa sectors.
One begins with several Yukawa matrices $\Gamma_i$ and $\Delta_i$, which usually have many free parameters,
multiply them by vevs $v_i$ and sum them to produce the mass matrices $M_d^0$ and $M_u^0$, see Eqs.~\eqref{Md0Mu0-general}.
After the bidiagonalization procedure \eqref{quark-rotations},
we obtain physical parameters $m_q$, $V_{\rm CKM}$,
as well as the quark coupling matrices $N_d$ and $N_u$, see Fig.~\ref{Fig:diagram}.

In general, the passage from $\Gamma_i$, $\Delta_i$ to $M_d^0$, $M_u^0$ is irreversible:
if one only knows $M_d^0$ and vevs, one cannot recover individual $\Gamma_i$.
Technically, it comes from the fact that $\Gamma_i$ are not
linearly independent. Indeed, in the general case, for each set of $v_i$,
it is possible to modify $\Gamma_i$ by some matrices $\delta \Gamma_i$ which sum up to the zero $3\times 3$ matrix:
$\sum_i v_i \,  \delta\Gamma_i = {\bf 0}_{3}$.
Thus, the Yukawa sectors based on $\Gamma_i$ and on $\Gamma_i + \delta\Gamma_i$ lead to the same mass matrix $M_d^0$.

Another problem is that, if matrices $\Gamma_i$, $\Delta_i$ are not generic, %exhibit some pattern within a specific multi-Higgs model,
it is not guaranteed that they can reproduce the known quark masses and mixing parameters at all, especially
when the vev alignment is also constrained by the scalar sector.
A nice illustration of this situation for the $A_4$ and $S_4$ symmetric 3HDMs can be found in \cite{GonzalezFelipe:2013xok,GonzalezFelipe:2014mcf}.

In the light of these difficulties, one is often forced to scan the multi-dimensional parameter space of the model
in a way which is intrinsically inefficient. If one knows vevs and randomly selects $\Gamma_i$, $\Delta_i$,
one obtains $M_d^0$ and $M_u^0$, which lead to quark masses and mixing very different
from their experimental values. One needs to repeat the scan many times,
trying to iteratively approach the measured values.

However, in certain classes of models one can invert the above procedure:
that is, knowing $M_d^0$ and vevs, one can uniquely reconstruct individual $\Gamma_i$.
This situation takes place, for example, if each $\Gamma_i$ lives in a different subspace
of the general $3\times 3$ complex matrix space.
In particular, in models with abelian symmetry groups with different scalars transforming
with different charges, the non-zero entries of the Yukawa matrices $\Gamma_i$ or $\Delta_i$ 
are non-overlapping and thus satisfy the above conditions.

The possibility of inversion significantly facilitates the phenomenological study of the model.
Instead of a random scan over $\Gamma_i$, $\Delta_i$, one takes the physical quark masses and mixing parameters as input,
parametrizes $V_{dL}$, $V_{dR}$, and $V_{uR}$ in a suitable way, and directly obtains a parameter space point
which automatically agrees with the experimental quark properties.
No parameter sets are wasted anymore.
Moreover, one can express the physical quark coupling matrices $N_d$ and $N_u$ via
quark masses and mixing as well as $V_{dL}$, $V_{dR}$, and $V_{uR}$.
In this way, one may derive certain predictions for FCNC 
or use some of their entries as
additional input parameters instead of $V_{dL}$, $V_{dR}$, $V_{uR}$.
Such a procedure will bring more control over FCNC within a given class of models.

The inversion procedure is available not only in models with natural flavor conservation, such as the Type-I or Type-II 2HDMs,
but also in the Branco-Grimus-Lavoura (BGL) two-Higgs-doublet model \cite{Branco:1996bq},
which allows for small FCNCs controlled by the third row of $V_{\rm CKM}$.
A similar inversion procedure was recently constructed in the $U(1)\times \Z_2$-symmetric 3HDM \cite{Das:2021oik},
where the quark sector closely resembled the BGL model.

\subsection{The inversion procedure in CP4 3HDM}

The first scan of the CP4 3HDM parameter space reported in \cite{Ferreira:2017tvy} was done in the traditional way,
by randomly choosing $\Gamma_i$, $\Delta_i$ and iteratively approaching the physical quark sector parameters.
In the CP4 3HDM, the matrices $\Gamma_2$ and $\Gamma_3$ (or $\Delta_2$ and $\Delta_3$) live in the same subspace,
so that, at the first glance, the inversion procedure is not expected to work.
However the entries of the two matrices are related by the $CP$ symmetry of order 4.
Thus, as already mentioned in \cite{Ferreira:2017tvy}, the inversion procedure does apply to the CP4 3HDM,
although it was not used in that scan.

In the present work, we develop this procedure for each case of the Yukawa sector \eqref{caseB1}--\eqref{caseB3}.
Starting from the physical quark parameters $m_q$, $V_{\rm CKM}$ and parametrizing the quark rotation matrices,
we are able to calculate $N_{d2}$, $N_{d3}$, $N_{u2}$, $N_{u3}$ for the physical quark couplings
with the second and third Higgs doublets $H_2^0$ and $H_3^0$ in the Higgs basis.
Although these scalars are not yet the mass eigenstates, the matrices provide a clear picture
of the FCNC magnitude and patterns to be expected in the model.

As we will show below, certain features of the matrices $N_{d2}$ and $N_{u2}$ are universal
for $B_1$, $B_2$, $B_3$ and closely resemble the BGL-like models or the $U(1)\times \Z_2$-symmetric 3HDM \cite{Das:2021oik}.
In particular, we will see that their FCNC couplings are controlled by the third row of the $V_{\rm CKM}$,
which makes them naturally small.
However there are also important differences with respect to the original BGL models
and the $U(1)\times \Z_2$-symmetric 3HDM.
One is that the symmetry transformation of our model is unavoidably mixes quark generations,
leading to FCNC patterns which do not always follow the familiar behavior.

Another difference with respect to the BGL-type models is that
the Yukawa matrices of CP4 3HDM are less constrained. As a result, 
the quark rotation matrices $V_{dL}$, $V_{dR}$, $V_{uL}$, $V_{uR}$,
do not, in general, exhibit the block-diagonal structure characteristic of the BGL models.

However, since the CP4 transformation acts on three generations by the irreducible representation decomposition $2+1$,
it automatically singles out one quark generation, which we assume to be the heaviest one.
As a result, we find it natural to explore matrices $V_{dL}$, $V_{dR}$, $V_{uL}$, $V_{uR}$
in the vicinity of the block-diagonal form.
With the inversion procedure implemented, we will study the generic magnitude and the patterns of the FCNC matrices
in the CP4 3HDM and check what is the smallest FCNC one can achieve for the physical values of $m_q$ and $V_{\rm CKM}$.

\subsection{Target values for the FCNC magnitude}\label{subsection-target}

How small do the off-diagonal elements of $N_d$ and $N_u$ need to be in order
to have a chance to satisfy the meson oscillation constraints?
The exact bounds depend not only on the masses of the new bosons
but also on other details of the scalar sector, such as their mixing angles,
partial cancellations among different scalars, as well as self-cancellation
between the scalar and pseudoscalar Yukawa couplings for each neutral Higgs boson \cite{Botella:2014ska,Nebot:2015wsa}.
For example, a scalar $S$ and a pseudoscalar $A$ of close masses
originating from the same neutral complex field $(S+iA)/\sqrt{2}$ coupled to quarks
tend to compensate each other's contribution to meson oscillation amplitude 
due to the relative $i^2=-1$ factor \cite{Botella:2014ska}.
In this work, we are {\em not} going to invoke these delicate mechanisms.
Instead, we address the question:
\begin{itemize}
\item[]
is it possible to achieve, within CP4 3HDM,
sufficiently small FCNC couplings which would satisfy all the neutral meson oscillation constraints
for a 1 TeV Higgs boson without relying on additional cancellation?
\end{itemize}
If we answer it in the affirmative, then there are good chances that a full phenomenological scan
of the scalar and Yukawa sectors of the CP4 3HDM, with all the cancellations included,
will identify viable points with reasonably heavy Higgs bosons.
Conversely, if finding such examples turns out impossible,
it means that the future phenomenological scan must either heavily rely of strong cancellations
or push all new Higgs bosons to multi-TeV range.
In this case, the chances to find a phenomenologically acceptable version 
of the explicitly CP4-invariant 3HDM will be bleak. 

Let us now define the target parameters.
Following \cite{Nebot:2015wsa}, we first rewrite the coupling matrix of a generic real scalar $S$ 
of unspecified $CP$ properties with down quarks as
\begin{equation}
\frac{1}{v}\bar{d}_{Li}\, (N_d)_{ij} d_{Rj} + h.c =
\bar d_{i}\left(A_{ij} + i B_{ij} \gamma^5\right) d_j\,,
\end{equation}
where
\begin{equation}\label{A-B}
A = \frac{N_d + N_d^\dagger}{2v}\,, \quad
i B = \frac{N_d - N_d^\dagger}{2v}\,.
\end{equation}
Both $A$ and $B$ are hermitean matrices.
For example, $K^0$--$\overline{K^0}$ oscillations place constraints on
$|a_{ds}| = |A_{12}|$ and $|b_{ds}| = |B_{12}|$, and so on.
Notice that, for a very asymmetric matrix $N_d$ exhibiting $|(N_{d})_{12}| \gg |(N_{d})_{21}|$,
we obtain $|a_{ds}| \approx |b_{ds}|$.
A similar construction for the up-quark sector allows us to constrain the $(uc)$ elements
with the aid of $D^0$--$\overline{D^0}$ oscillations.

We consider the off-diagonal FCNC elements acceptable for a 1 TeV scalar
if they satisfy the following upper limits borrowed from \cite{Nebot:2015wsa}:
% %%%%%%%   this is for 1 TeV %%%%%%%%%%
\begin{subequations}\label{oscillation-constraints}
\begin{align}
K^0 - \overline{K^0}: &\qquad |a_{ds}| < 3.7\times 10^{-4}\,, \quad |b_{ds}| < 1.1\times 10^{-4}\,,\label{constraint-ds}\\
B^0 - \overline{B^0}: &\qquad |a_{db}| < 9.0\times 10^{-4}\,, \quad |b_{db}| < 3.4\times 10^{-4}\,,\label{constraint-db}\\
B_s^0 - \overline{B_s^0}: &\qquad |a_{sb}| < 45\times 10^{-4}\,, \quad\  |b_{sb}| < 17\times 10^{-4}\,,\label{constraint-sb}\\
D^0 - \overline{D^0}: &\qquad |a_{uc}| < 5.0\times 10^{-4}\,, \quad |b_{uc}| < 1.8\times 10^{-4}\,.\label{constraint-uc}
\end{align}
\end{subequations}
% %%%%%%%   this is for 250 GeV %%%%%%%%%%
%\begin{subequations}\label{oscillation-constraints}
%\begin{align}
%K^0 - \overline{K^0}: &\qquad |a_{ds}| < 0.9\times 10^{-4}\,, \quad |b_{ds}| < 0.3\times 10^{-4}\,,\label{constraint-ds}\\
%B^0 - \overline{B^0}: &\qquad |a_{db}| < 2.3\times 10^{-4}\,, \quad |b_{db}| < 0.8\times 10^{-4}\,,\label{constraint-db}\\
%B_s^0 - \overline{B_s^0}: &\qquad |a_{sb}| < 11\times 10^{-4}\,, \quad\  |b_{sb}| < 4.2\times 10^{-4}\,,\label{constraint-sb}\\
%D^0 - \overline{D^0}: &\qquad |a_{uc}| < 1.3\times 10^{-4}\,, \quad |b_{uc}| < 0.45\times 10^{-4}\,.\label{constraint-uc}
%\end{align}
%\end{subequations}
Notice that in all cases, the upper bounds on $|b|$ are comparable with the Cheng-Sher Ansatz
$\sqrt{m_im_j}/v$, which would give $0.9\times 10^{-4}$, $6\times 10^{-4}$, $25\times 10^{-4}$, $2\times 10^{-4}$, respectively.
For a lower mass of the scalar $S$, the upper limits on these couplings will decrease proportionally.

We stress once more that these constraints are not to be taken as strict pass-or-fail bounds.
They just indicate the typical FCNC values which, for a 1 TeV scalar, could be compatible 
with meson oscillations without resorting to additional cancellation mechanism.
However models violating these constraints by a large factor will most likely fail
the meson oscillation test even with possible cancellation effects included.

Below, we will check one by one all seven Yukawa sectors given 
in Eqs.~\eqref{cases-first-group} or \eqref{cases-second-group}.
If a sector manages to produce ``viable'' points, that is, parameter sets
which pass all the above constraints, we consider this case promising, 
generate a sample of these viable points 
and will include it in a future phenomenological study.
If a sector is unable to produce points passing simultaneously all the constraints,
we say the this sector is ``ruled out''.

%%%%%%%%%%%%%%%%%%%%%%%%%%%%%%%%%%%%%%%%%%%%%%%%%%%%%%%%%%%%%%%

\section{FCNCs in the CP4 3HDM: general expressions}\label{section-FCNC-cases}

\subsection{Case $B_1$}

Using the expressions for matrices $\Gamma_i$ in Eq.~\eqref{caseB1} and the vevs
parametrized via $\beta$ and $\psi$ in Eq.~\eqref{vevs}, we can explicitly compute $M_d^0$ as well as
$N_{d2}^0$ and $N_{d3}^0$. Our goal is to relate them, that is, to express $N_{d2}^0$ and $N_{d3}^0$
via $M_d^0$, which will then allow us to express $N_{d2}$ and $N_{d3}$ in terms of quark masses and quark rotation matrices.

We begin with $N_{d2}^0$, Eq.~\eqref{Nd20Nd30-general}, which can be expressed as
\begin{equation}
N_{d2}^0 = M_d^0 \cot\beta - \frac{v}{\sqrt{2} s_\beta} \Gamma_1 = R_3^0 \cdot M_d^0\,.
\end{equation}
Here,
\begin{equation}
R_3^0 = \mmmatrix{\cot\beta}{0}{0}{0}{\cot\beta}{0}{0}{0}{-\tan\beta} = \cot\beta \left({\bf 1}_3 - \frac{P_3}{c_\beta^2} \right),\label{R30}
\end{equation}
with ${\bf 1}_3$ being the unit $3\times 3$ matrix and $P_3 = {\rm diag}(0, 0, 1)$ being the projector on axis $3$.
After the mass matrix bidiagonalization, we get
\begin{equation}
N_{d2} = V_{dL}^\dagger N_{d2}^0 V_{dR} = V_{dL}^\dagger R_3^0 V_{dL} \cdot V_{dL}^\dagger M_d^0 V_{dR} = R_3 \cdot M_d\,.
\end{equation}
This leads to a simple expression for the FCNC matrix $N_{d2}$, which makes it clear that
the off-diagonal elements of $N_{d2}$ are fully controlled by the third row of the matrix $V_{dL}$:
\begin{equation}
(N_{d2})_{ij} = \cot\beta\, m_{d_j}\delta_{ij} - \frac{m_{d_j}}{c_\beta s_\beta} (V_{dL, 3i})^* V_{dL, 3j}\,.\label{B1-Nd2-general}
\end{equation}
A similar analysis holds for the up-quark sector, leading to
\begin{equation}
(N_{u2})_{ij} = \cot\beta\, m_{u_j}\delta_{ij} - \frac{m_{u_j}}{c_\beta s_\beta} (V_{uL, 3i})^* V_{uL, 3j}\,.\label{B1-Nu2-general}
\end{equation}
These expressions are familiar from the BGL model and the $U(1)\times \Z_2$-symmetric 3HDM studied in \cite{Das:2021oik}.
This is not surprising: with our definition of the Higgs basis, the second doublet $H_2$ of our model
matches the second doublet of the 2HDM in the Higgs basis.

For $N_{d3}$, which has no counterpart in the BGL model, we obtain
\begin{equation}
N_{d3}^0 = \frac{1}{s_\beta} \mmmatrix{-M_{22}^*}{-M_{21}^*}{-M_{23}^*}{M_{12}^*}{M_{11}^*}{M_{13}^*}{0}{0}{0}\,,\label{B1-Nd30}
\end{equation}
where $M_{ij}$ stand for the elements of the mass matrix $M_d^0$.
The properties of the CP4 transformation lead to the unusual feature of this expression, namely, that 
the vev alignment angle $\psi$ disappears at the price of using $(M_d^0)^*$ instead of $M_d^0$.
Next, one can represent $N_{d3}^0$ as
\begin{equation}
N_{d3}^0 = \frac{1}{s_\beta} P_4 \cdot M_d^{0 *} \cdot R_2\,, \quad
\mbox{where}\quad P_4 = \mmmatrix{0}{-1}{0}{1}{0}{0}{0}{0}{0}\,, \quad
R_2 = \mmmatrix{0}{1}{0}{1}{0}{0}{0}{0}{1}\,. \label{B1-Nd30-general}
\end{equation}
The subscripts in these matrices indicate that $P_4$ is a projector onto the subspace $(1,2)$
together with an order-4 rotation and $R_2$ is just a reflection within this subspace.
After the quark field rotations, we get
\begin{equation}
N_{d3} = V_{dL}^\dagger N_{d3}^0 V_{dR} =
\frac{1}{s_\beta} V_{dL}^\dagger P_4 V_{dL}^*\cdot V_{dL}^T M_d^{0 *}  V_{dR}^*\cdot V_{dR}^T R_2 V_{dR}
= \frac{1}{s_\beta} P_4^{(dL)} \cdot M_d \cdot R_2^{(dR)}\,. \label{B1-Nd3-general}
\end{equation}
Here, we used the fact that the diagonal matrix $M_d$ is real. We also defined
\begin{equation}
P_4^{(dL)} = V_{dL}^\dagger P_4 V_{dL}^*\,, \quad
R_2^{(dR)} = V_{dR}^T \, R_2 \, V_{dR}\,.\label{P4R2}
\end{equation}
The matrix $P_4^{(dL)}$ can be recast in a more revealing form.
First we write $(P_4)_{ij} = - \epsilon_{ij3}$, where epsilon is the fully antisymmetric invariant tensor satisfying
\begin{equation}
\epsilon_{ijk} = V_{ii'} V_{jj'} V_{kk'}\cdot \epsilon_{i'j'k'} \cdot (\det V)^*\label{epsilon-ijk}
\end{equation}
for any unitary matrix $V\in U(3)$.
In particular, it holds for $V = V_{dL}$.
Now, understanding $V$ as $V_{dL}$, we can write $P_4$ as
\begin{eqnarray}
(P_4^{(dL)})_{ij} &=& V^\dagger_{im} \left(- \epsilon_{mn3}\right) V^*_{nj}
%= - V^\dagger_{im} V^\dagger_{jn}\, \epsilon_{mn3}\nonumber\\
= - V^\dagger_{im} V^\dagger_{jn} \cdot V_{mm'} V_{nn'} V_{3k} \cdot \epsilon_{m'n'k}\, (\det V)^*\nonumber\\
% &=& - \delta_{im'}\, \delta_{jn'} \cdot \epsilon_{m'n'k} V_{3k}\, \det V^* \nonumber\\
&=& - \epsilon_{ijk} V_{3k} \, (\det V)^*\,.\label{P4-caseB1}
\end{eqnarray}
Thus, $P_4^{(dL)}$ is also constructed with the aid of the same third row of $V_{dL}$. Finally,
\begin{equation}
(N_{d3})_{ij} = - \frac{1}{s_\beta}\epsilon_{ikm} V_{dL, 3m}\, m_{d_k}\, (R_2^{(dR)})_{kj}\, \, (\det V_{dL})^*\,,
\label{B1-Nd3-general2}
\end{equation}
where
$(R_2^{(dR)})_{kj} = V_{dR,1k} V_{dR,2j} + V_{dR,2k} V_{dR,1j} + V_{dR,3k} V_{dR,3j}$.
A similar expression holds for $N_{u3}$
\begin{equation}
(N_{u3})_{ij} = - \frac{1}{s_\beta}\epsilon_{ikm} V_{uL, 3m}\, m_{u_k}\, (R_2^{(uR)})_{kj}\, \, (\det V_{uL})^*\,,
\label{B1-Nu3-general}
\end{equation}
with
$(R_2^{(uR)})_{kj} = V_{uR,1k} V_{uR,2j} + V_{uR,2k} V_{uR,1j} + V_{uR,3k} V_{uR,3j}$.
In summary, using the quark masses and $V_{\rm CKM}$ as input and assuming any form of
$V_{dL}$, $V_{dR}$, $V_{uL}$, $V_{uR}$ (of course, satisfying $V_{uL}^\dagger V_{dL} = V_{\rm CKM}$),
one can directly compute $N_{d2}$, $N_{d3}$, $N_{u2}$, $N_{u3}$
through Eqs.~\eqref{B1-Nd2-general}, \eqref{B1-Nd3-general2}, \eqref{B1-Nu2-general}, and \eqref{B1-Nu3-general}.

\subsection{Case $B_2$}

The Yukawa matrices for case $B_2$, Eq.~\eqref{caseB2}, are similar to case $B_1$, with the role of rows and columns interchanged.
The analysis proceeds in a similar way, with $V_{dL}$ and $V_{dR}$ being exchanged.
\begin{equation}
N_{d2}^0 = M_d^0 \cot\beta - \frac{v}{\sqrt{2} s_\beta} \Gamma_1 = M_d^0 \cdot R_3^0\,,
\end{equation}
with the same $R_3^0$ as in Eq.~\eqref{R30}. After the quark field rotations, we obtain
\begin{equation}
N_{d2, ij} = \cot\beta\, m_{d_i}\delta_{ij} - \frac{m_{d_i}}{c_\beta s_\beta} (V_{dR, 3i})^* V_{dR, 3j}\,.\label{B2-Nd2-general}
\end{equation}
A similar expression holds for the up-quark sector:
\begin{equation}
N_{u2, ij} = \cot\beta\, m_{u_i}\delta_{ij} - \frac{m_{u_i}}{c_\beta s_\beta} (V_{uR, 3i})^* V_{uR, 3j}\,.\label{B2-Nu2-general}
\end{equation}
The differences with respect to case $B_1$ are: $m_{d_i}$ instead of $m_{d_j}$ and $V_{dR}$ instead of $V_{dL}$
(with the corresponding differences for the up sector).
Switching from the left to the right-handed fields is important:
$V_{dR}$ and $V_{uR}$ can be chosen independently from each other,
while $V_{dL}$ and $V_{uL}$ used in case $B_1$ were not independent.

The expression for $N_{d3}^0$ resembles Eq.~\eqref{B1-Nd30}:
\begin{equation}
N_{d3}^0 = \frac{1}{s_\beta} \mmmatrix{M_{22}^*}{-M_{21}^*}{0}{M_{12}^*}{-M_{11}^*}{0}{M_{32}^*}{-M_{31}^*}{0}
= \frac{1}{s_\beta} R_2 \cdot M_d^{0 *} \cdot P_4\,,\label{B2-Nd30}
\end{equation}
with the same $P_4$ and $R_2$ as in \eqref{B1-Nd30-general}.
After the quark field rotations, we get
\begin{equation}
N_{d3} = V_{dL}^\dagger N_{d3}^0 V_{dR} =
\frac{1}{s_\beta} V_{dL}^\dagger R_2 V_{dL}^*\cdot V_{dL}^T M_d^{0 *}  V_{dR}^*\cdot V_{dR}^T P_4 V_{dR}
= \frac{1}{s_\beta} R_2^{(dL)} \cdot M_d \cdot P_4^{(dR)}\,. \label{B2-Nd3-general}
\end{equation}
Here, as before, we defined
\begin{eqnarray}
(R_2^{(dL)})_{ij} &=& \left(V_{dL}^\dagger \, R_2 \, V_{dL}^*\right)_{ij}
= \left(V_{dL,1i} V_{dL,2j} + V_{dL,2i} V_{dL,1j} + V_{dL,3i} V_{dL,3j}\right)^*\,,\nonumber\\
(P_4^{(dR)})_{ij} &=& \left(V_{dR}^T \, P_4 \, V_{dR}\right)_{ij} = - \epsilon_{ijk} (V_{dR,3k})^*\, \det V_{dR}\,.
\label{B2-P4R2}
\end{eqnarray}
For the up-quark sector, we have
\begin{eqnarray}
N_{u3} &=& \frac{1}{s_\beta} R_2^{(uL)} \cdot M_u \cdot P_4^{(uR)}\,, \label{B2-Nu3-general}\\
(R_2^{(uL)})_{ij} &=& \left(V_{uL}^\dagger \, R_2 \, V_{uL}^*\right)_{ij}
= \left(V_{uL,1i} V_{uL,2j} + V_{uL,2i} V_{uL,1j} + V_{uL,3i} V_{uL,3j}\right)^*\,,\nonumber\\
(P_4^{(uR)})_{ij} &=& \left(V_{uR}^T \, P_4 \, V_{uR}\right)_{ij} = - \epsilon_{ijk} (V_{uR,3k})^*\, \det V_{uR}\,.
\end{eqnarray}

\subsection{Case $B_3$}

The structure of Yukawa matrices in case $B_3$ is different, Eq.~\eqref{caseB3}.
Let us begin with $N_{d3}^0$:
\begin{equation}
N_{d3}^0 = \frac{1}{s_\beta} \mmmatrix{0}{0}{-M_{23}^*}{0}{0}{M_{13}^*}{M_{32}^*}{-M_{31}^*}{0}\,,\label{B3-Nd30}
\end{equation}
which can be expressed as
\begin{equation}
N_{d3}^0 = \frac{1}{s_\beta} \left(P_4 M_{d}^{0 *} P_3 + P_3 M_{d}^{0 *} P_4\right)
\end{equation}
using the same matrix $P_4$ as before and the same projector on the third axis $P_3 = {\rm diag}(0, 0, 1)$.
A similar expression holds for the up sector.
After the quark field rotations, we get
\begin{eqnarray}
N_{d3} &=& V_{dL}^\dagger N_{d3}^0 V_{dR} =
\frac{1}{s_\beta} \left(P_4^{(dL)} M_d P_3^{(dR)} + P_3^{(dL)} M_d P_4^{(dR)}\right)\,,\label{B3-Nd3-general}\\
N_{u3} &=& V_{uL}^\dagger N_{u3}^0 V_{uR} =
\frac{1}{s_\beta} \left(P_4^{(uL)} M_u P_3^{(uR)} + P_3^{(uL)} M_u P_4^{(uR)}\right)\,.\label{B3-Nu3-general}
\end{eqnarray}
As before, for any matrix $A$, we understand $A^{L}$ as $V_{L}^\dagger \, A \, V_{L}^*$
and  $A^{R} = V_{R}^T \, A \, V_{R}$; for example, $(P_3^{(uR)})_{ij} = V_{uR,3i} V_{uR,3j}$.

For $N_{d2}$, we can construct two different presentations thanks to the properties of the matrix $\Gamma_1$:
\begin{equation}
N_{d2}^0 =
M_d^0 \cot\beta - \frac{1}{s_\beta c_\beta}\mmmatrix{M_{11}}{M_{12}}{0}{M_{21}}{M_{22}}{0}{0}{0}{M_{33}}
= M_d^0 \cot\beta - \frac{1}{s_\beta c_\beta}\mmmatrix{M_{22}^*}{-M_{21}^*}{0}{-M_{12}^*}{M_{11}^*}{0}{0}{0}{M_{33}^*}\,.
\label{B3-Nd20}
\end{equation}
The latter form can be compactly written as
\begin{equation}
N_{d2}^0 = M_d^0 \cot\beta - \frac{1}{s_\beta c_\beta} \left[P_3 M_d^{0 *} P_3 - P_4 M_d^{0 *} P_4\right]\,,
\end{equation}
which, with the aid of the same matrices as in \eqref{B3-Nd3-general}, leads to
\begin{equation}
N_{d2} = M_d \cot\beta - \frac{1}{s_\beta c_\beta} \left(P_3^{(dL)} M_d P_3^{(dR)} - P_4^{(dL)} M_d P_4^{(dR)}\right)\,,
\label{B3-Nd2-second}
\end{equation}
A similar expression holds for the up sector:
\begin{eqnarray}
N_{u2} &=&M_u \cot\beta - \frac{1}{s_\beta c_\beta} \left(P_3^{(uL)} M_u P_3^{(uR)} - P_4^{(uL)} M_u P_4^{(uR)}\right)\,.
\label{B3-Nu2-general}
\end{eqnarray}

\section{Limiting FCNC: a qualitative analysis}\label{section-minimal}

\subsection{A toy model}

In order to see how small the FCNCs can in principle be, let us begin with a toy version
of the CP4 3HDM, in which the true CKM matrix is replaced by a simplified block-diagonal form
parametrized with the single Cabibbo angle $\theta_C$.
Let us also assume that all four quark rotation matrices $V_{dL}, V_{dR}, V_{uL}, V_{uR}$,
which are input parameters within the inversion procedure,
can also be chosen in the same block diagonal form:
\begin{equation}
V_{dL}, V_{dR}, V_{uL}, V_{uR} \sim \mmmatrix{\times}{\times}{0}{\times}{\times}{0}{0}{0}{\times}
= e^{i\delta}\mmmatrix{c_\theta e^{i\alpha}}{s_\theta e^{i\zeta}}{0}%
{-s_\theta e^{-i\zeta}}{c_\theta e^{-i\alpha}}{0}%
{0}{0}{e^{i\gamma}}\label{param-block}\,.
\end{equation}
The parameters $\theta$, $\alpha$, $\zeta$, $\delta$, $\gamma$ in each matrix are independent, 
subject only to $V_{\rm CKM} = V_{uL}^\dagger V_{dL}$. %The determinant of this matrix is $\exp(3i\delta + i\gamma)$.
Under these assumptions, the expressions for the matrices $N$ are dramatically simplified.

Let us begin with case $B_1$. The matrices $N_{d2}, N_{u2}$ become diagonal:
\begin{equation}
N_{d2} = \mmmatrix{m_d \cot\beta }{0}{0}{0}{m_s\cot\beta }{0}{0}{0}{-m_b \tan\beta }\,,\quad
N_{u2} = \mmmatrix{m_u \cot\beta }{0}{0}{0}{m_c\cot\beta }{0}{0}{0}{-m_t \tan\beta }\,,\label{B1-Nd2-Nu2-toy}
\end{equation}
which is well known from the original BGL model.
As for $N_{d3}, N_{u3}$, we first notice that $P_4^{(dL)} = e^{-2i\delta_{dL}} P_4$, while
\begin{equation}
R_2^{(dR)} = e^{2i\delta}\mmmatrix{-s_{2\theta}e^{i(\alpha-\zeta)}}{c_{2\theta}}{0}{c_{2\theta}}{s_{2\theta}e^{-i(\alpha-\zeta)}}{0}{0}{0}{e^{2i\gamma}}\,,
\end{equation}
where all parameters correspond to the matrix $V_{dR}$.
As a result, we obtain
\begin{equation}
N_{d3} = \frac{e^{2i(\delta_{dR}-\delta_{dL})}}{s_\beta} \mmmatrix{-m_s c_{2\theta}}{-m_s s_{2\theta}e^{-i(\alpha-\zeta)}}{0}%
{-m_d s_{2\theta}e^{i(\alpha-\zeta)}}{m_d c_{2\theta}}{0}{0}{0}{0}\,,
\label{B1-Nd3-toy}
\end{equation}
where all unlabeled angles refer to $V_{dR}$.
A similar expression holds for the up sector:
\begin{equation}
N_{u3} = \frac{e^{2i(\delta_{uR}-\delta_{uL})}}{s_\beta} \mmmatrix{-m_c c_{2\theta}}{-m_c s_{2\theta}e^{-i(\alpha-\zeta)}}{0}%
{-m_u s_{2\theta}e^{i(\alpha-\zeta)}}{m_u c_{2\theta}}{0}{0}{0}{0}\,,
\label{B1-Nu3-toy}
\end{equation}
with the unlabeled angles corresponding to $V_{uR}$.

We observe that, in this toy model, none of the neutral Higgses generates any FCNC
involving the third generation quarks $b$ and $t$,
thus easily satisfying the constraints from $B$/$B_s$-meson oscillations.
FCNCs only appear between the first two generations and lead to
\begin{eqnarray}
&& |a_{ds}| \approx |b_{ds}| \approx \frac{m_s}{2v}\frac{\sin 2\theta_{dR}}{\sin\beta}
= 2\times 10^{-4} \, \frac{\sin 2\theta_{dR}}{\sin\beta}\,,\label{B1-estimate-K}\\
&& |a_{uc}| \approx |b_{uc}| \approx \frac{m_c}{2v}\frac{\sin 2\theta_{uR}}{\sin\beta}
= 25\times 10^{-4} \, \frac{\sin 2\theta_{uR}}{\sin\beta}\,.\label{B1-estimate-D}
\end{eqnarray}
Comparing them with Eqs.~\eqref{constraint-ds},~\eqref{constraint-uc}, we conclude that
we need $\theta_{dR} < 0.2$ and $\theta_{uR} < 0.03$ to satisfy the meson oscillation constraints.
%generic values of $\theta_{dR}$, $\theta_{uR}$ are insufficient.
Since these two angles are independent free parameters,
it is technically possible to satisfy the constraints
and even to eliminate FCNCs altogether by setting $\theta_{dR} = \theta_{uR} = 0$.
Then, the only non-trivial feature of the matrices $N$'s is the peculiar pattern
of their diagonal values: for example, $H_{3}^0$ couples to $\bar u u$ proportionally to $m_c$,
not $m_u$.

Within the toy model, case $B_2$ leads to exactly the same diagonal matrices $N_{d2}$ and $N_{u2}$ as in Eq.~\eqref{B1-Nd2-Nu2-toy}.
The matrices $N_{d3}$ and $N_{u3}$ have the same form as in \eqref{B1-Nd3-toy} and \eqref{B1-Nu3-toy} up to transposition;
the convention now is that the unlabeled angles correspond to $V_{dL}$, $V_{uL}$.
However, these two matrices are no longer independent.
In particular, it is impossible to select both $\theta_{dL} = \theta_{uL} = 0$ since they are related
with via a sizable Cabibbo angle $|\theta_{dL} +\theta_{uL}| \ge  \theta_C \approx 0.22$.
Incidentally, the upper values on $\theta_{dL}$ and $\theta_{uL}$ computed for a 1 TeV scalar
are barely compatible with $\theta_C$.
Another way out is to select case $A$ at least for one of the two sectors,
which would eliminate FCNCs in that sector completely
and would allow to keep the other angle small.

Interestingly, case $B_3$ turns out to be incompatible with the toy model.
Indeed, limiting the quark rotation matrices to the block-diagonal form
forces us to either set $\Gamma_2 = \Gamma_3 = 0$ or to assume $\sin\beta = 0$.
In either case, one would observe that $M_d^0 \propto \Gamma_1$, which
would force $m_d = m_s$, in conflict with experiment.

\subsection{Implications for realistic models}\label{subsection-minimal-realistic}

The lesson we draw from the toy model is that it is possible to suppress
the FCNC couplings involving $b$ quark even for the realistic $V_{\rm CKM}$.
To achieve this, we should keep all the quark rotation matrices
as close as possible to the block-diagonal form \eqref{param-block}.

To get some insight, let us choose case $(B_1,B_1)$ and suppose that $V_{uR}$, $V_{dR}$, and $V_{uL}$ are block-diagonal
and the compute $V_{dL} = V_{uL} V_{\rm CKM}$.
Then $N_{u2}$ keeps the diagonal form as in Eq.~\eqref{B1-Nd2-Nu2-toy},
while $N_{u3}$ is still given by the same expression \eqref{B1-Nu3-toy} as in the toy model.
In the down-quark sector, the third row of $V_{dL}$ is proportional to the third
row of the CKM matrix: $V_{dL, 3i} = V_{uL, 33} V_{{\rm CKM}, 3i}$.
For $N_{d2}$ we can use the general expression \eqref{B1-Nd2-general} but replace $V_{dL, 3i}$ with $V_{{\rm CKM}, 3i}$.
In this way one obtains for $N_{d2}$ exactly the same FCNC Ansatz as in the BGL model,
which is driven in our case by CP4.
As for $N_{d3}$, we return to the general expression \eqref{B1-Nd3-general2}.
This matrix is not of the block-diagonal form anymore, but, when evaluating its third column, we encounter
$m_{d_k} (R_2^{(dR)})_{k3} = m_{b} \delta_{k3} (V_{dR,33})^2$.
As a result, the third {\em column} of $N_{d3}$
\begin{equation}
(N_{d3})_{i3} = -\frac{1}{s_\beta} \epsilon_{i3m} V_{uL, 33} V_{{\rm CKM}, 3m} m_{b} (V_{dR,33})^2\, (\det V_{dL})^*
\end{equation}
can be expressed via the third {\em row} of $V_{\rm CKM}$ in the following peculiar way:
\begin{equation}
(N_{d3})_{i3} \simeq \frac{m_b}{s_\beta} \triplet{V_{ts}}{-V_{td}}{0}\,.\label{B1-min-Nd3}
\end{equation}
Here, the symbol $\simeq$ indicates that we omitted the inessential phase factor.
It is interesting to note that, up to this phase factor,
this column is independent of the quark rotation angles and, due to $|V_{td}| \ll |V_{ts}|$,
the largest FCNCs involving $b$-quarks come
in the form of $b\to d$, not $b \to s$ transition.
Finally, all other elements of the matrix $(N_{d3})_{i3}$ can include either $m_d$ or $m_s$ but not $m_b$.
Indeed, the term with $m_b$ corresponding to $k=3$ in Eq.~\eqref{B1-Nd3-general2}
forces $j=3$, which we just considered.

%%%%%%%%%%%%%%%%%%%%%%%%%%%%%%%%%%%%%%%%%%%%%%%%%%%%%%%%%%%%%%%%%%%%%%%%%%%%

\section{Numerical results}\label{section-numerical}

Equipped with the general expressions and guided by the insights from the toy model,
we are ready to address the main question formulated in Section~\ref{subsection-target}:
is it possible to find examples of the CP4 3HDM Yukawa sectors
which satisfy all the meson oscillation constraints given in Eqs.~\eqref{oscillation-constraints}?

The first step of our numerical study is to implement the inversion procedure
described in Section~\ref{section-controlling}.
We start with the experimentally known quark masses and the CKM matrix,
for which we take the following values:
\begin{eqnarray}
&& 
(m_d, m_s, m_b) [\mbox{GeV}] = (0.0047,\, 0.096,\, 4.18)\,, \quad
(m_u, m_c, m_t) [\mbox{GeV}] = (0.0022,\, 1.28,\, 173.1)\,,\nonumber\\
&&\sin\theta_{12} = 0.22496\,, \quad 
\sin\theta_{13} = 0.003617\,, \quad 
\sin\theta_{23} = 0.04165\,, \quad 
\sin\delta = 0.949\,.\label{mq-CKM}
\end{eqnarray}
These are not the latest measurements; 
we just borrowed the valued used in \cite{Ferreira:2017tvy} to facilitate the comparison.
Next, we choose the quark rotation matrices in such a way that, in the original basis
before the quark mass matrices diagonalization, we recover $M_d^0$ and $M_u^0$
exactly of the type that is required for each Yukawa sector.
For cases $A$, $B_1$, and $B_2$, this can be done analytically,
while for case $B_3$ we resort to a numerical procedure.
Details of this step for each of the four cases are described in the Appendix.

The angles and phases of the quark rotation matrices represent the space
in which we perform numerical scan.
This scan can done in different ways.
What we label below as a ``full scan'' corresponds to a random selection,
with the uniform probability distribution,
of all the rotation angles and phases within their full ranges.
In a ``restricted scan'', we try to choose the quark rotation matrices as close as possible
to the block-diagonal form \eqref{param-block}.
We do this by choosing the angles $\theta_{13}$ and $\theta_{23}$
from the interval $[-\theta_{max}, \theta_{max}]$, where $\theta_{max}$ is some pre-defined value,
such as $\pi/100$.

In all plots, unless specified otherwise, the default value of the vev ratio parameter $\tan\beta = 1$.
The other vev alignment angle, $\psi$, does not appear in matrices $N$'s.

\subsection{Case $(B_1,B_1)$}

\subsubsection{Kaon and $D$-meson constraints}

\begin{figure}[H]
	\centering
	\includegraphics[width=0.48\textwidth]{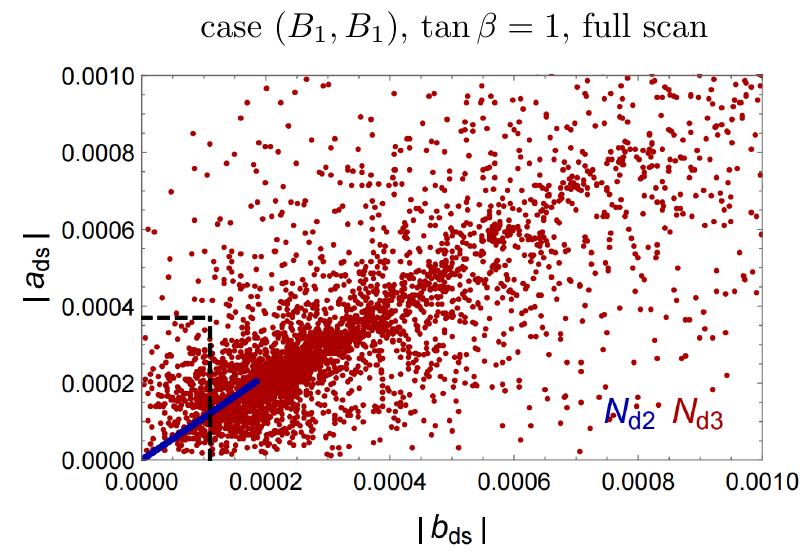}
	\includegraphics[width=0.48\textwidth]{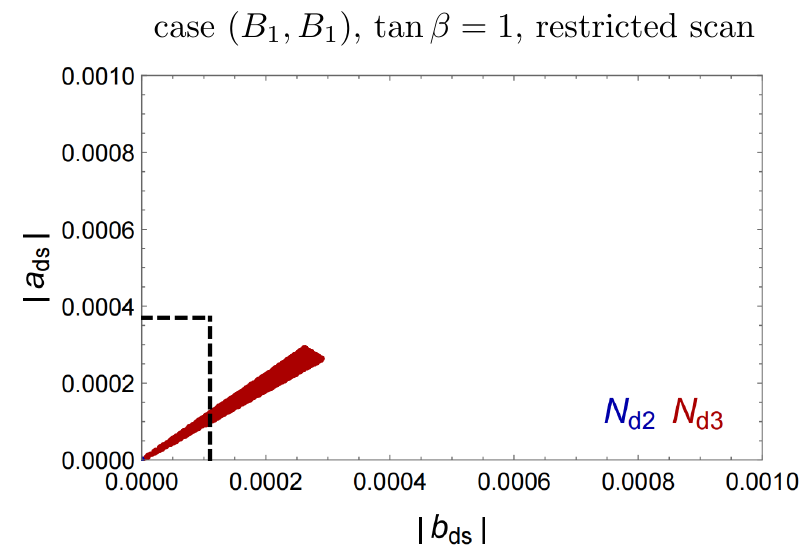}
	\caption{The impact of kaon oscillations. Shown are the values of $|a_{ds}|$, $|b_{ds}|$
	obtained in a full scan (left) and a restricted scan with $\theta_{max} = \pi/100$ (right).
	The dashed box shows the limits Eq.~\eqref{constraint-ds}.}
	\label{fig-B1B1-K}
\end{figure}

We begin our numerical exploration with a detailed analysis of case $(B_1,B_1)$.
We present in Fig.~\ref{fig-B1B1-K} the impact of the kaon oscillation constraints on the FCNC couplings.
The scatter plot shows the values of $|a_{ds}|$ vs. $|b_{ds}|$
obtained in a full scan (left) and a restricted scan with $\theta_{max} = \pi/100$ (right).
On both scatter plots, we show separately the values coming from $N_{d2}$ (blue) and $N_{d3}$ (red).

As we see, even in a full scan, we obtain
many points which pass the kaon oscillation constraints Eq.~\eqref{constraint-ds}, indicated by the dashed box,
for both $N_{d2}$ and $N_{d3}$.
Restricting the scan to nearly block-diagonal quark matrices further suppresses FCNCs
(in the right plot, the blue points corresponding to $N_{d2}$ shrink at zero).
This is in accordance with the toy model expectations,
where $N_{d2}$ becomes a diagonal matrix, Eq.~\eqref{B1-Nd2-Nu2-toy},
while $N_{d3}$ approaches the block-diagonal form \eqref{B1-Nd3-toy}.
Due to $(N_{d3})_{21} \gg (N_{d3})_{12}$, we obtain $|a_{ds}|\approx |b_{ds}|$, which explains the straight line segment shape of the plot.
The upper limit of the last plot ($N_{d3}$ for small $\theta_{max}$)
is explained by the estimate \eqref{B1-estimate-K}, taking into account that $\sin\beta = 1/\sqrt{2}$.

In short, the kaon oscillations constraints can be easily satisfied within case $(B_1,B_1)$.

\begin{figure}[H]
	\centering
	\includegraphics[width=0.48\textwidth]{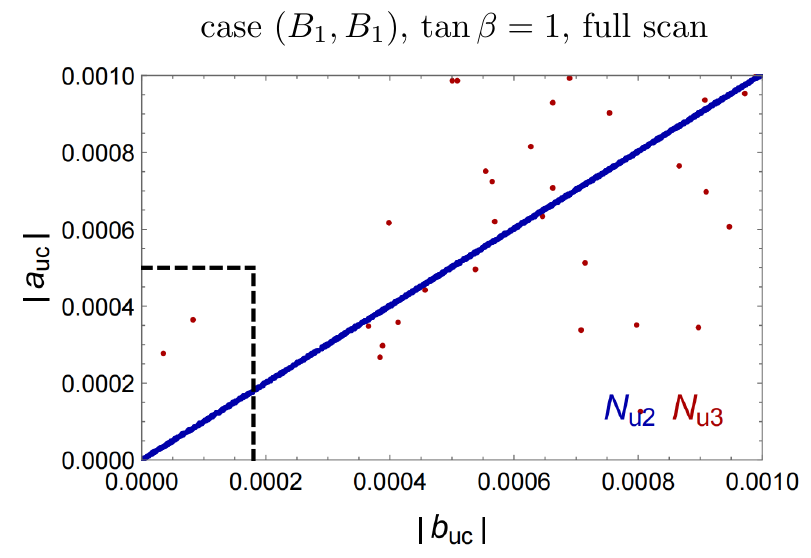}
	\includegraphics[width=0.48\textwidth]{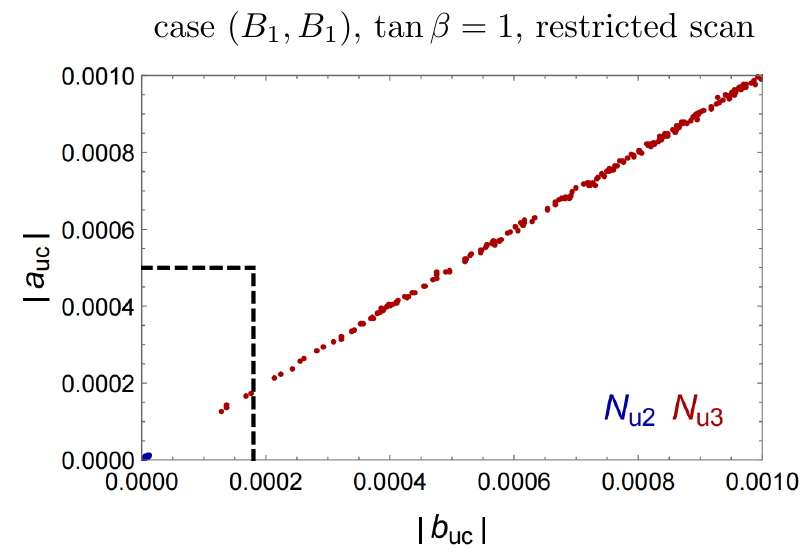}
	\caption{The impact of $D$-meson oscillations.
	Shown are the values of $|a_{uc}|$, $|b_{uc}|$ in a full scan (left) 
	and a restricted scan with $\theta_{max} = \pi/100$ (right).
	The dashed box shows the limits Eq.~\eqref{constraint-uc}.}
	\label{fig-B1B1-D}
\end{figure}

In Fig.~\ref{fig-B1B1-D}, we explore the FCNC effects in the up-quark matrices and compare them
with the $D$-meson oscillation constraints Eq.~\eqref{constraint-uc}.
The overall picture here is the same as for kaons,
with the exception that only a few points fall inside the box of the allowed $|a_{uc}|$, $|b_{uc}|$ values.
This is not surprising: the estimate \eqref{B1-estimate-D} shows that we need a rather small $\theta_{uR}$
to pass the $D$-meson oscillation constraints.
Unfortunately, with the procedure we use for building case $B_1$, we cannot fully control the value of this angle.
However we checked that the points inside the box indeed correspond to $|\theta_{uR}| < 0.025$.
Thus, the $D$-meson oscillations constraints can also be satisfied.

\begin{figure}[H]
	\centering
	\includegraphics[width=0.5\textwidth]{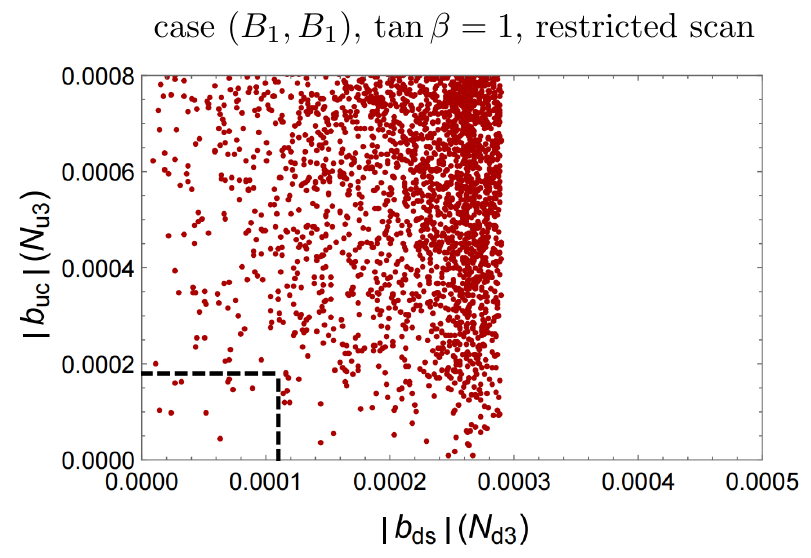}
	\caption{The values of $|b_{uc}|$, $|b_{ds}|$ obtained from $N_{u3}$, $N_{d3}$
	in a restricted scan with $\theta_{max} = \pi/100$.}
	\label{fig-B1B1-K-D}
\end{figure}

In Fig.~\ref{fig-B1B1-K-D} we show that the kaon and $D$-meson constraints can also be satisfied simultaneously.
Here, we only show the most challenging case: $|b_{uc}|$ coming from $N_{u3}$ and $|b_{ds}|$ coming from $N_{d3}$.
To increase the density of points, we run the scan with $10^5$ points instead of 5000 points used in the previous plots.
We see that such points do exist and correspond to small $\theta_{uR}$ and $\theta_{dR}$.
We also checked that increasing $\tan\beta$ allows for a mild additional suppression
of the FCNC couplings and further increases the number of points simultaneously passing the two sets of constraints.

\subsubsection{$B$/$B_s$-meson constraints}

\begin{figure}[H]
	\centering
	\includegraphics[width=0.48\textwidth]{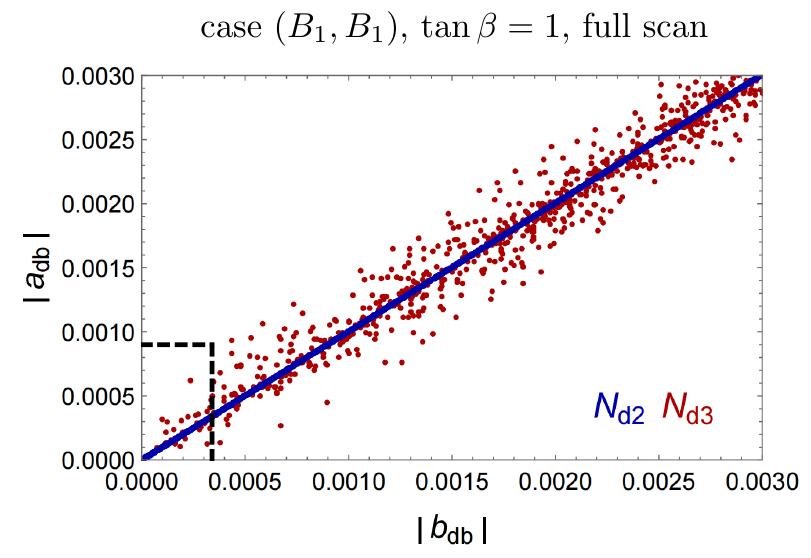}
	\includegraphics[width=0.48\textwidth]{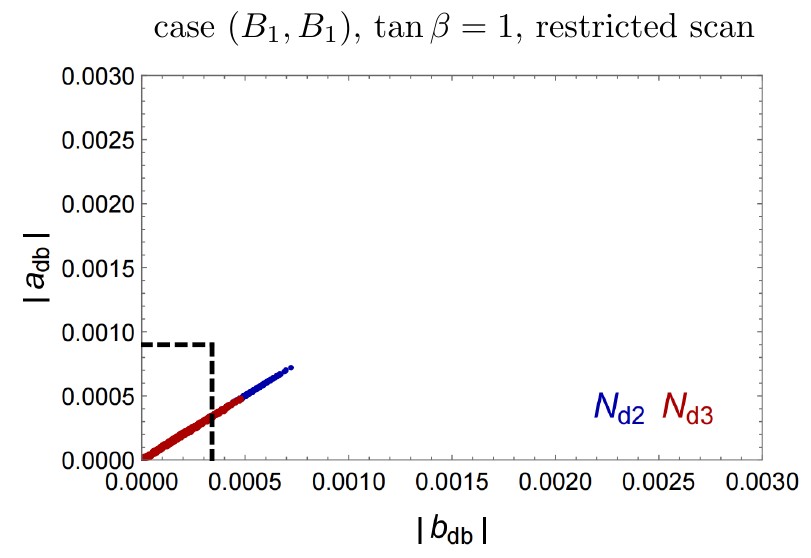}
	\caption{The impact of $B$-meson oscillations.
	Shown are the values of $|a_{db}|$, $|b_{db}|$ in a full scan (left) and a restricted scan with $\theta_{max} = \pi/100$ (right).
	The dashed box shows the limits Eq.~\eqref{constraint-db}.}
	\label{fig-B1B1-B}
\end{figure}

Next, we look into the $B$-meson oscillation constraints.
Fig.~\ref{fig-B1B1-B} shows the same sequence of plots for values $|a_{db}|$, $|b_{db}|$
together with the $B$-meson constraints \eqref{constraint-db}.
We also observed a similar picture (not shown) for the $B_s$ mesons, constraints being less tight.
As expected, small $\theta_{max}$ brings the matrices close to the block-diagonal form
and, as a result, suppresses the $b$-quark FCNCs.
Thus, $B$-mesons do not represent an obstacle once we keep the quark rotation matrices close to the block-diagonal form.
In fact, we could find points which satisfy the $B$-meson oscillation
constraints for scalar masses as low as 200 GeV.

\subsubsection{Case $(B_1,B_1)$: the overall picture}

The above numerical results lead to the following overall situation for case $(B_1,B_1)$.
\begin{itemize}
\item
The off-diagonal elements of the Higgs-quark coupling matrices $N_{d2}$, $N_{d3}$, $N_{u2}$, $N_{u3}$
can be controlled through appropriate parameters of the quark rotation matrices.
\item
Different parameters control different FCNC couplings. $B$/$B_s$-meson oscillation constraints
for a 1 TeV scalar are easily satisfied by choosing the quark rotation matrices sufficiently close to the block-diagonal form.
The kaon and $D$-meson constraints can be satisfied by the restricting the mixing angle $\theta_{12}$
in the matrices $V_{dR}$ and $V_{uR}$.
\item
$D$-meson oscillations place the strongest constraint for a generic scan.
They must be included in a viable phenomenological analysis.
Since the previous analysis \cite{Ferreira:2017tvy} did not include them,
all the $(B_1,B_1)$ points considered there as viable would, mostly likely,
be ruled out by the $D$-meson constraints. 
\end{itemize}

\begin{figure}[H]
	\centering
	\includegraphics[width=0.48\textwidth]{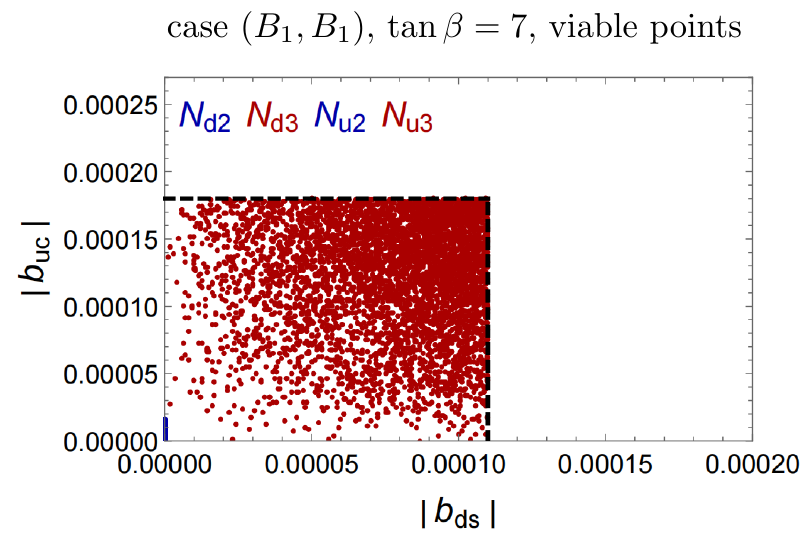}
	\includegraphics[width=0.48\textwidth]{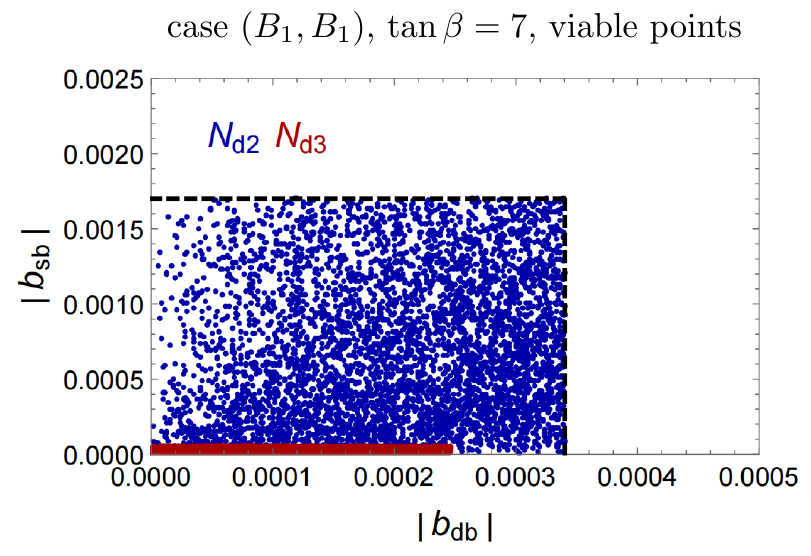}
	\caption{Constraints from the kaon and $D$-meson oscillations (left) and from $B$/$B_s$-meson oscillations (right)
	for the subset of points which satisfy all the meson oscillation constraints.
	}
	\label{fig-B1B1-good}
\end{figure}

Finally, we could also select a subset of points which pass the constraints from all four neutral meson systems.
In Fig.~\ref{fig-B1B1-good} we show how these ``viable'' points are distributed on the $D$-meson vs kaon plane (left)
and on the $B_s$ vs $B$-meson plane (right). This run corresponds to $\tan\beta = 7$;
increasing the value of $\tan\beta$ allowed us to get more viable points with respect to the previous scans.
Curiously, the plots show that couplings of $H_2^0$ and $H_3^0$ to quarks are shaped by different 
meson oscillation constraints. Indeed, $N_{d2}$ and $N_{u2}$ (blue points)
satisfy the kaon and $D$-meson constraints by large margin,
and their main limitations come from $B$ physics.
For $N_{d3}$ and $N_{u3}$ (red points), we observe the opposite trend: $B$ physics constraints play a minor role,
and the strongest restrictions come from kaons and $D$-mesons.
This is a manifestation of the structurally different forms of the corresponding matrices.
Also, from the distribution of these points inside the boxes we see that some points are compatible
with the Higgs bosons with masses of a few hundred GeV.
In a future work, we plan to couple them with a scalar sector scan and build viable benchmark models
based on $(B_1,B_1)$ Yukawa sector of the CP4 3HDM.

\subsection{Cases $(B_1,B_3)$, $(B_3,B_1)$, and $(B_3,B_3)$}

Case $B_3$, defined by the Yukawa matrices \eqref{caseB3}, is a peculiar one.
As already mentioned, this case does not possess a limit that could correspond to the toy model
considered in Section~\ref{section-minimal}.
Therefore, diagonalizing a realistic case $B_3$ mass matrix must involve all three rotation angles.
As a result, the $(ds)$ and $(uc)$ entries of the matrices $N$ will receive
contributions proportional to the third generation quark masses.
This unavoidable three-generation mixing is especially important in the up-quark sector.
The large top quark mass gives sizable contributions to $(N_{u2})_{12}$ and $(N_{u2})_{21}$
and leads to an unacceptably large $D$-meson oscillation amplitude.

\begin{figure}[H]
	\centering
	\includegraphics[width=0.48\textwidth]{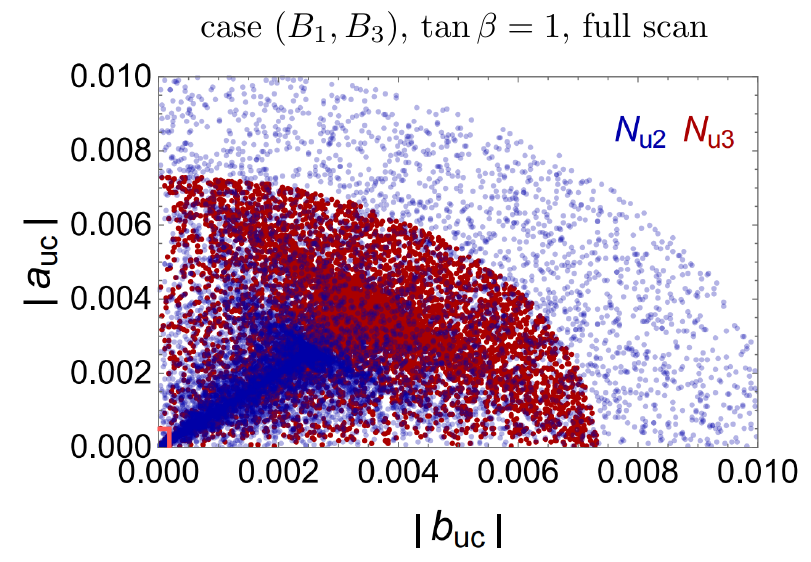}
	\includegraphics[width=0.48\textwidth]{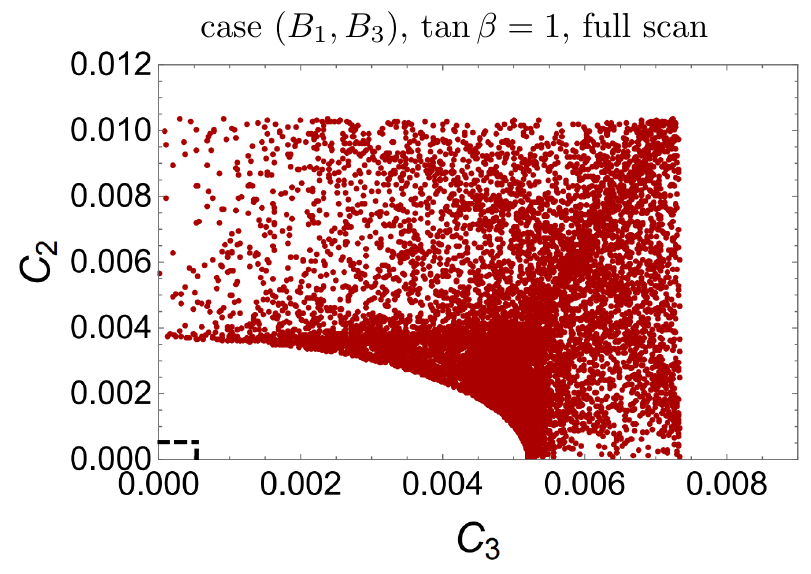}
	\caption{Left: the values of $|a_{uc}|$ vs. $|b_{uc}|$ from $N_{u2}$ and $N_{u3}$
	in a full scan of case $(B_1,B_3)$.
	Right: the interplay between $N_{u2}$ and $N_{u3}$ expressed in terms
	of combined parameters $c_2$ and $c_3$ defined in Eq.~\eqref{c2c3}.}
	\label{fig-B1B3-D}
\end{figure}

Consider, for example the Yukawa model $(B_1, B_3)$,
in which the down-quark sector displays distributions similar to what we just studied.
In Fig.~\ref{fig-B1B3-D}, left, we show the usual plot of $|a_{uc}|$ and $|b_{uc}|$ obtained from
$N_{u2}$ and $N_{u3}$ in a full scan of this model.
The points cover a sizable area extending up to values of $0.01$.
The box with the $D$-meson oscillation constraints \eqref{constraint-uc}
is barely visible on this plot, but we checked that there exist points
with either $N_{u2}$ or $N_{u3}$ falling inside the box.
However there are no examples in which both $N_{u2}$ and $N_{u3}$ pass the constraints.
This is illustrated in Fig.~\ref{fig-B1B3-D}, right, where we compare $D$-mixing FCNC contributions
from $N_{u2}$ and $N_{u3}$ in terms of the following combined quantities:
\begin{equation}
c_2 \equiv \sqrt{|a_{uc}|^2+|b_{uc}|^2}\quad \mbox{from $N_{u2}$}\,, \quad
c_3 \equiv \sqrt{|a_{uc}|^2+|b_{uc}|^2}\quad \mbox{from $N_{u3}$}\,.
\label{c2c3}
\end{equation}
This plot was computed for $\tan\beta=1$;
for other values of $\tan\beta$, it has the same shape
but is stretched along one of the axes.
We observe a sharp lower bound on a quadratic combination for $c_2$ and $c_3$,
which depends on $\tan\beta$ and which is always much bigger than
the experimental constraints.

In short, we could not find any point for the model $(B_1,B_3)$ which would come
sufficiently close to satisfying the $D$-meson oscillation constraints.
We conclude that CP4 3HDM with Yukawa sector of type $(B_1,B_3)$
cannot produce viable points.
The scan reported in \cite{Ferreira:2017tvy} claimed to find suitable parameter space points
in case $(B_1,B_3)$ but those points would be ruled out by the $D$-meson oscillation constraints.

We also ruled out case $(B_3,B_3)$ on the same grounds;
predictions for the $D$-meson oscillations are a way off from the experimental constraints.

\begin{figure}[h]
	\centering
	\includegraphics[width=0.48\textwidth]{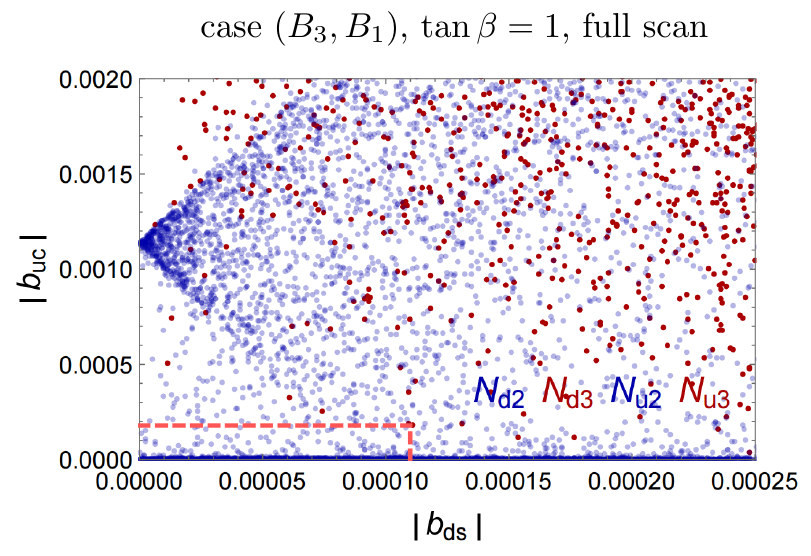}
	\includegraphics[width=0.48\textwidth]{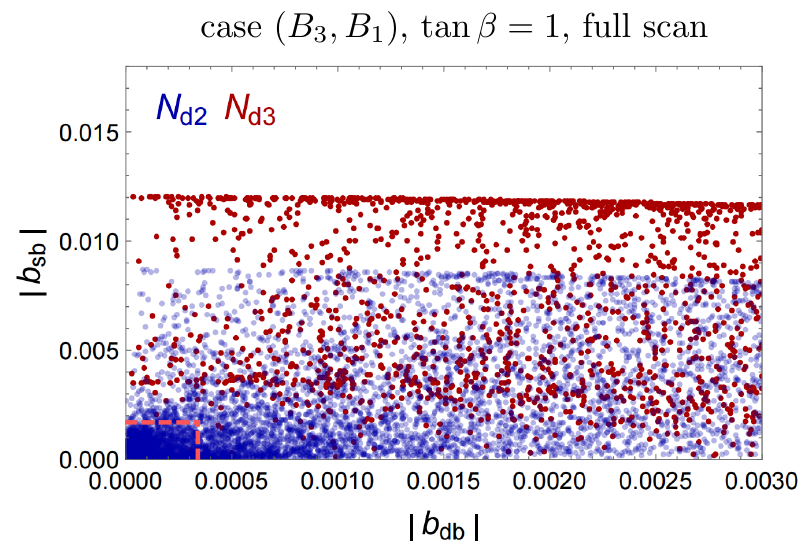}
	\caption{Constraints from the kaon and $D$-meson oscillations (left) and from $B$/$B_s$-meson oscillations (right)
	on the FCNC quantities in case $(B_3,B_1)$.}
	\label{fig-B3B1}
\end{figure}

For the model $(B_3,B_1)$, too, finding points which would pass all the meson oscillation constraints turned out impossible.
The plots in Fig.~\ref{fig-B3B1} illustrate the problems we had.
The left plot shows that, in stark constraint with case $(B_1,B_1)$ given in Fig.~\eqref{fig-B1B1-good}, it is extremely difficult 
to find a point which would pass constraints from kaons and $D$-mesons simultaneously, especially for $H_3^0$.
The right plot of Fig.~\ref{fig-B3B1} shows comparison of $B$ and $B_s$ constraints; 
it is again $N_{d3}$ which falls outside the box.
With all meson constraints combined, we find that every point violates at least one constraint by a factor of few.

\subsection{Cases $(A,B_2)$, $(B_2,A)$, and $(B_2,B_2)$}

The Yukawa sector of case $A$ given by \eqref{caseA} seems to be the ideal choice
if we look to eliminate FCNCs altogether.
However we cannot assume case $A$ for both up and down quarks simultaneously,
as this combination cannot produce the $CP$-violating entries of the CKM matrix,
see Appendix, Section~\ref{appendix-caseA}.
Thus, the safest choice is case $(A,B_2)$, in which we avoid all the kaon and $B$/$B_s$ constraints
and need to care only about $D$-mesons.

\begin{figure}[H]
	\centering
	\includegraphics[width=0.5\textwidth]{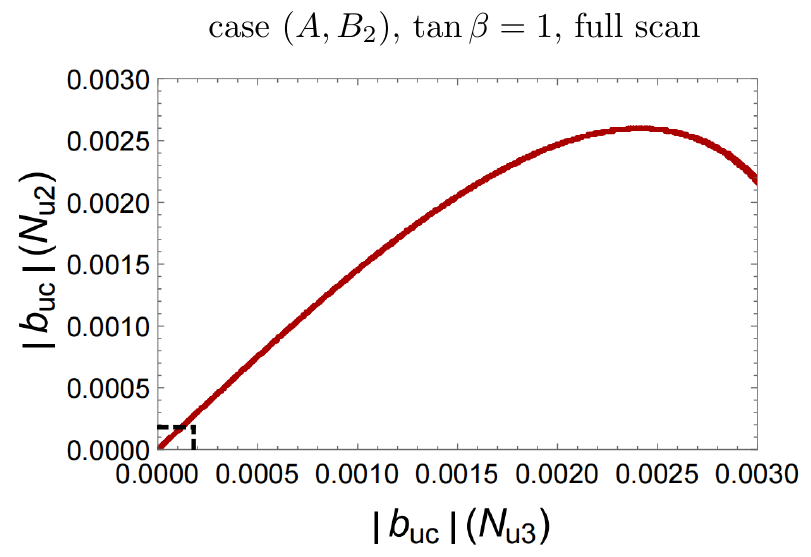}
	\caption{The interplay between the values of $|b_{uc}|$ calculated from $N_{u2}$ and $N_{u3}$ in a full scan
	of case $(A,B_2)$.}
	\label{fig-AB2-D}
\end{figure}

Since case $B_2$ admits the toy model limit, we expect to have many parameter space points
with up-quark rotation matrices $V_{uL}$, $V_{uR}$ close to the block-diagonal form.
Within the procedure we use for case $B_2$, described in Appendix, Section~\ref{appendix-caseB2},
we cannot choose arbitrary rotation angles.
However when performing a scan, we observed that $|a_{uc}| \approx |b_{uc}|$
for both $N_{u2}$ and $N_{u3}$ and that many points indeed satisfy the $D$-meson constraints.
This is illustrated by Fig.~\ref{fig-AB2-D}, where we show the values of
$|b_{uc}|$ calculated from $N_{u2}$ and $N_{u3}$.
We could find points with the FCNC contributions significantly smaller 
than the $D$-meson constraints Eq.~\eqref{constraint-uc}.
Thus, case $(A,B_2)$ offers many viable points.
In a future analysis, we plan to update the procedure which could parametrically 
suppress the $(uc)$ couplings in $N_{u2}$ and $N_{u3}$. 

In case $(B_2,A)$, we have the opposite situation: the $D$-meson constraints become irrelevant
as there are no FCNCs in the up sector, while there arise problems for kaons and $B$-physics.
We found that the kaon oscillation constraints do not represent a severe problem;
many points passing them for $N_{d2}$ and $N_{d3}$ could be found.
However the combined analysis of $B$ and $B_s$-meson represents a serious obstacle.

\begin{figure}[H]
	\centering
	\includegraphics[width=0.5\textwidth]{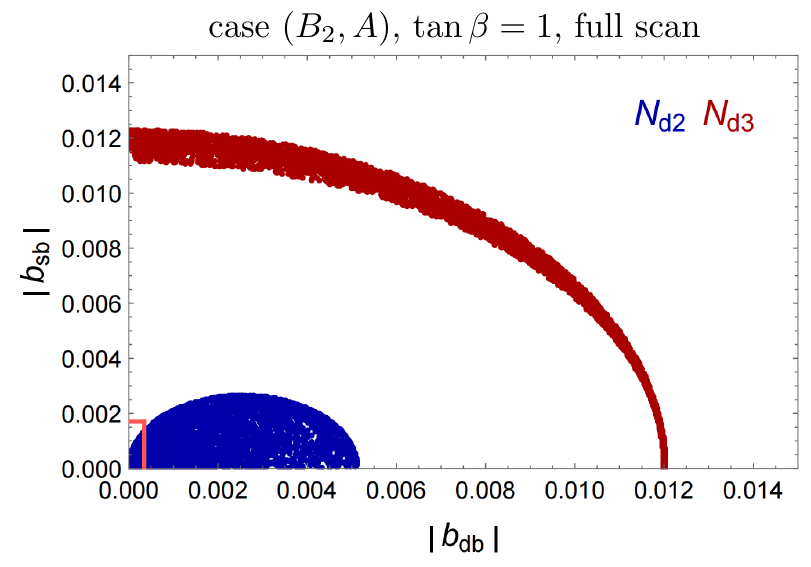}
	\caption{The values of $|b_{db}|$ and $|b_{sb}|$ for $N_{d2}$ and $N_{d3}$ in a full scan
	of case $(B_2,A)$.}
	\label{fig-B2A-B-Bs}
\end{figure}

In Fig.~\ref{fig-B2A-B-Bs}, we show a simultaneous check of the $B$ and $B_s$ constraints on a full scan.
We see that the FCNC couplings from $N_{d2}$ are rather suppressed and can pass the constraints
but $N_{d3}$ fails by far. This check alone rules out case $(B_2,A)$.

\begin{figure}[H]
	\centering
	\includegraphics[width=0.48\textwidth]{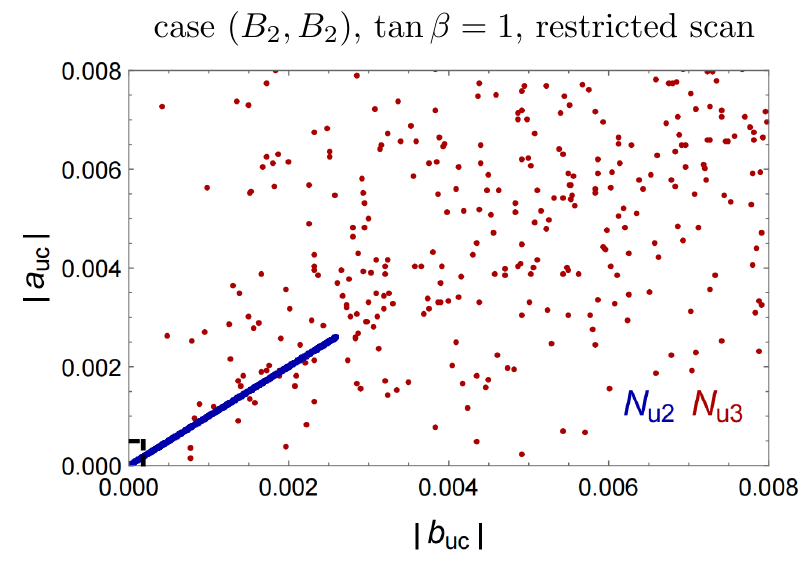}
	\includegraphics[width=0.48\textwidth]{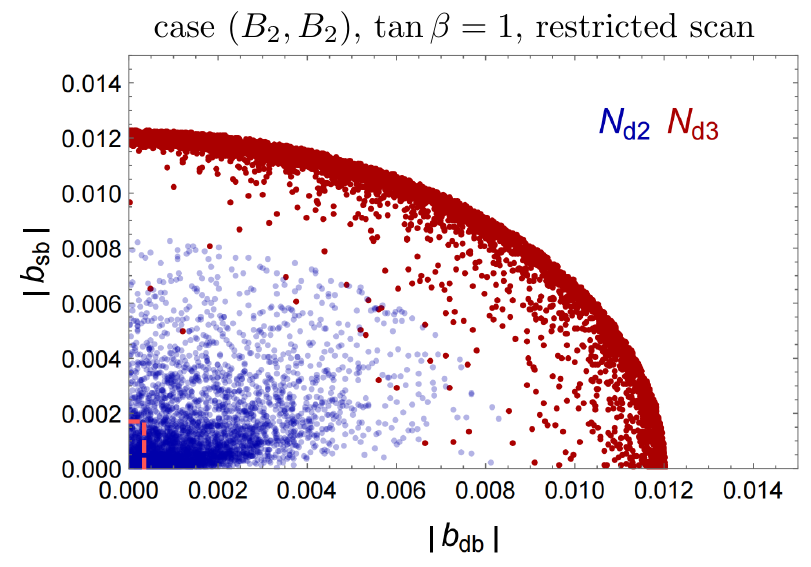}
	\caption{Constraints from the $D$-meson (left) and the $B$/$B_s$-meson (right) oscillations
	on the restricted scan with $\theta_{max} = \pi/100$ within case $(B_2,B_2)$.}
	\label{fig-B2B2}
\end{figure}

Finally, case $(B_2,B_2)$ leads to problems both in the up and down-quark sectors.
Just as in the previous case, the constraints coming from kaon oscillations can be satisfied
in a restricted scan with $\theta_{max} = \pi/100$.
However, the $D$-meson constraints are nearly impossible to satisfy for $N_{u3}$ even 
within a restricted scan with $\theta_{max}=\pi/100$, see Fig.~\ref{fig-B2B2}, left.
Moreover, when one tries to suppress $D$-meson mixing by restricting the ranges of rotation angles,
the matrix $N_{d3}$ runs into a conflict with constraints for $B$ or $B_s$-mesons,
as shown in Fig.~\ref{fig-B2B2}, right. This, too, is similar to case $(B_2,A)$.
We conclude that case $(B_2,B_2)$ is ruled out by the combination 
of the $D$-meson and $B$/$B_s$-meson oscillation constraints.

It is interesting to highlight once again an important difference between
cases $(B_1,B_1)$ and $(B_2,B_2)$, which we already discussed in Section~\ref{section-minimal}.
For $(B_1,B_1)$, working in the vicinity of the block-diagonal form of quark rotation matrices,
we could simultaneously suppress the kaon and $D$-meson FCNC contributions.
This was possible because those elements were governed by the right-handed quark rotation matrices
$V_{dR}$ and $V_{uR}$, which are independent from each other.
As a result, we could generate many viable points in case $(B_1,B_1)$.
For $(B_2,B_2)$, these FCNC elements are controlled by the left-handed matrices
$V_{dL}$ and $V_{uL}$, which are not independent.
This leads to the clash between down-sector and up-sector
FCNC matrices $N_{d3}$ and $N_{u3}$ and makes it impossible to find viable points.

\section{Discussion and conclusions}\label{section-discussion}

Neutral meson oscillations put to severe test all multi-Higgs models
that feature Higgs-induced tree-level flavor changing neutral couplings \cite{Sher:2022aaa}.
CP4 3HDM, the three-Higgs-doublet model based on the exotic $CP$ symmetry of order 4,
does contain FCNCs with peculiar CP4-driven patterns.
Nevertheless, the first phenomenological scan of the model presented in \cite{Ferreira:2017tvy}
identified parameter space points which pass many experimental constraints
including kaon and $B$/$B_s$-meson oscillations.
Later, these parameter space points were ruled out by the light charged Higgs searches \cite{Ivanov:2021pnr},
but the intrigue remained. 
Can the FCNC effects be strongly suppressed within CP4 3HDM? 
In which CP4 Yukawa sectors can it happen and what controls the magnitude of FCNCs? 
Are there viable examples of the model which pass all the neutral meson oscillation constraints
for the new Higgs bosons not heavier than, say, 1 TeV?

In this paper, we systematically investigated all these questions.
Following \cite{Ferreira:2017tvy}, we work in the scalar alignment regime of the CP4 3HDM,
which guarantees that the 125 GeV Higgs couplings to quarks are as in the SM, 
and that the FCNCs can arise only from the heavy new scalars.

First, we constructed the inversion procedure for CP4-invariant Yukawa sectors.
That is, we showed how to recover the parameters of the model starting
from the physical observables, quark masses and the CKM matrix,
and using appropriately shaped quark rotation matrices.
This inversion procedure offers a huge advantage over \cite{Ferreira:2017tvy} 
for a phenomenological scan of the entire parameter space of the model 
because, by construction, each point of the scan agrees with 
the experimental values of the quark properties.

Our second result is the collection of analytic expressions for the 
matrices $N_{d2}$, $N_{d3}$ and $N_{u2}$, $N_{u3}$ which describe
how the new Higgses from the second and third doublets (in the Higgs basis we use)
couple to the down and up-quarks.
The off-diagonal entries of these matrices are the FCNC contributions which we want to suppress.
These matrices are written in terms of physical quark parameters 
and quark rotation matrix elements, which offers clear insights 
which parameters control which FCNC elements.

Third, we numerically studied whether the FCNC couplings can simultaneously pass the constraints
imposed by oscillations in the four neutral meson systems: kaons, $D$-mesons, $B$-mesons, and $B_s$-mesons.
Since the contributions depend on the new scalars masses, we computed the effects for 
the reference mass of 1 TeV. 

In our study, we did not rely on any possible cancellation among several Higgs contributions
to meson oscillations.
We did this on purpose: we wanted to check whether the FCNC couplings of all neutral scalars can be simultaneously
suppressed within any CP4 3HDM scenario without any help of an additional cancellation mechanism.
Examples of such mechanisms do exist \cite{Botella:2014ska,Nebot:2015wsa} but their effect is not robust
as it depends on the details of the scalar sector.
If a particular benchmark model leads to destructive interference of Higgs constributions,
then the FCNC constraints could be satisfied with scalars masses below 1 TeV.
Whether this indeed happens in specific models will become clear after the full phenomenological scan of the model
is done, which is delegated to a future work.

In total, there are eight possible CP4-invariant Yukawa sectors. 
It turns out that most of them lead to FCNC contributions 
to meson oscillations which cannot be simultaneously kept small 
in the four neutral meson systems.
Often, this comes from structural features of the Yukawa matrices,
so that no choice of the free parameters could bring all FCNCs under control. 

We find that, out of the eight CP4 3HDM Yukawa sectors, only two benchmark scenarios 
are capable of producing viable parameter space points passing all four meson constraints.
\begin{itemize}
\item 
Benchmark scenario $(A,B_2)$, in which the down-quark sector is completely free from FCNC.
The only constraints arise from the $D$-meson oscillations and can be easily satisfied
as the magnitude of FCNCs can be parametrically suppressed.
\item 
Benchmark scenario $(B_1,B_1)$, in which both up and down-quark sectors exhibit
FCNCs but their magnitudes are small if the quark rotation
matrices are close to the block-diagonal form.
\end{itemize}
In both cases, we generated large samples of viable points.
All other scenarios fail at least one of the meson constraint by a large margin; 
we found that they could become viable only for new Higgses as heavy as several TeV.

We also noticed the very important role of the $D$-meson oscillation constraints,
which are sufficient to rule out entire scenarios of the CP4 3HDM Yukawa sectors
irrespective of their free parameters.
Notice that the previous scan \cite{Ferreira:2017tvy} did not include the $D$-meson check.
We suspect that all the parameter space points found there would be ruled out by $D$-mesons.

Our results naturally lead to several follow-up studies.
Focusing on either of the two benchmark scenarios,
one can now perform a full phenomenological scan of the parameter space of the CP4 3HDM,
including the scalar sector. Due to the intrinsic interplay between the scalar and Yukawa sectors,
it will be a non-trivial exercise to satisfy simultaneously the experimental constraints 
on the new scalars and on the fermions. 
If such benchmark points are found, 
they can be used to make further collider predictions.

One can also include the lepton sector of the CP4 3HDM. A neutrino mass model
based on CP4 was described in \cite{Ivanov:2017bdx}, but it relied on the unbroken CP4 symmetry.
Whether spontaneously broken CP4 can lead to an acceptable charged lepton and neutrino sectors
remains an open question.

In summary, the CP4 3HDM achieves remarkably much for a multi-Higgs model based on a single symmetry,
and, as we show here, it can survive all the neutral meson oscillation constraints.
There remains much to be studied in this exotic but viable model, which will be the subject of the forthcoming papers.

\section*{Acknowledgments}

This work was supported by the National Natural Science Foundation of China
(Grant No. 11975320) and the Fundamental Research Funds for the Central Universities, Sun Yat-sen University.
R.P.~is supported in part by the Swedish Research Council grant, contract number 2016-05996, 
as well as by the European Research Council (ERC) under the European Union's Horizon 2020 research 
and innovation programme (grant agreement No 668679).

\appendix

\section{Choosing quark rotation matrices}

As explained in Section~\ref{subsection-inversion}, we rely on the inversion procedure to render the numerical scan efficient.
We begin with the physical quark masses and the CKM matrix and try to find the rotation matrices
$V_{dL}$, $V_{uL}$, $V_{dR}$, $V_{uR}$, which lead to the mass matrices $M_d^0$ and $M_u^0$
computed as
\begin{equation}
M_d^0 = V_{dL} D_d V_{dR}^\dagger\,,\quad
M_u^0 = V_{uL} D_u V_{uR}^\dagger
\end{equation}
of the form required for each case $A$, $B_1$, $B_2$, $B_3$.
This requirement constrains our choice of the quark rotation matrices in a non-trivial way. 
Here, we describe how we tackle this problem.

\subsection{Case $B_1$}
Let us begin with case $B_1$. Suppose we are given a generic set of vevs $v_i$.
Then, as it is seen from matrices $\Gamma_i$ in Eq.~\eqref{caseB1},
the first two rows of $M_d^0$ are filled with generic complex entries.
However the third row, which comes from $\Gamma_1$ alone,
is conditioned by the following two constraints:
\begin{equation}
\mbox{case $B_1$:}\quad (M_d^0)_{32} = [(M_d^0)_{31}]^*\,, \quad (M_d^0)_{33} \ \mbox{is real}.
\label{M-conditions-B1}
\end{equation}
In order to meet these requirements in our inversion approach, we employ a two-step procedure.
First, we begin with the diagonal down-type quark mass matrix $D_d$ and choose {\em generic} unitary
$V^{(gen.)}_{dL}$ and $V^{(gen.)}_{dR}$. The resulting matrix $M_d^{(gen.)}$ does not satisfy \eqref{M-conditions-B1},
but we numerically evaluate it at this stage.
At the second step, we {\em additionally} rotate the right-handed fields by the correcting matrix $C_{dR}$ to get
\begin{equation}
M_d^0 = M_d^{(gen.)}\, C_{dR}\,, \quad \mbox{where}\quad
C_{dR} = \mmmatrix{c_\theta e^{i\alpha}}{s_\theta}{0}{-s_\theta}{c_\theta e^{-i\alpha}}{0}{0}{0}{e^{i\xi_3}}\,.
\end{equation}
If we denote the third row of $M_d^{(gen.)}$ as $(\tilde g_{31},\, \tilde g_{32},\, \tilde g_{33})$, all of them being complex,
then the conditions in Eqs.~\eqref{M-conditions-B1} imply:
\begin{equation}
\tilde g_{31}^* s_\theta + \tilde g_{32}^* c_\theta e^{i\alpha} =
\tilde g_{31} c_\theta e^{i\alpha} - \tilde g_{32} s_\theta \,, \quad
\tilde g_{33} e^{i\xi_3}\ \mbox{is real}.
\end{equation}
The solution always exists and corresponds to
\begin{equation}
\tan\theta = \frac{|\tilde g_{31} - \tilde g_{32}^*|}{|\tilde g_{31}^* + \tilde g_{32}|}\,,\quad
\alpha = \arg(\tilde g_{31}^* + \tilde g_{32}) - \arg(\tilde g_{31} - \tilde g_{32}^*) = \arg[(\tilde g_{31}^*)^2 - (\tilde g_{32})^2]\,,
\end{equation}
together with $\xi_3 = - \arg \tilde g_{33}$.
If we have case $B_1$ for both up-quark and down-quark sectors,
we can select $V^{(gen.)}_{dL}$ as we wish, compute $V_{uL}$, and, at the first step,
choose arbitrary $V^{(gen.)}_{dR}$ and $V^{(gen.)}_{uR}$.
After evaluating $M_d^{(gen.)}$ and $M_u^{(gen.)}$, we find the parameters of the correcting matrices $C_{dR}$ and  $C_{uR}$.
Since we do not correct the left-handed fields, the CKM matrix is preserved.
Thus, we arrive at a viable model with the quark rotation matrices $V_{dL} = V^{(gen.)}_{dL}$
and $V_{dR} = C_{dR}^\dagger V^{(gen.)}_{dR}$.

\subsection{Case $B_2$}\label{appendix-caseB2}

Case $B_2$ mirrors the previous case; the two conditions to be satisfied involve elements of the third column of matrix $M_d^0$:
\begin{equation}
\mbox{case $B_2$:}\quad (M_d^0)_{23} = [(M_d^0)_{13}]^*\,, \quad (M_d^0)_{33} \ \mbox{is real}.
\label{M-conditions-B2}
\end{equation}
Clearly, they can be met through the same two-step procedure, which involves, at the second step,
a correcting matrix in left-handed field space.
The problem is that, in this case, the relation between $V_{dL}$ and $V_{uL}$ will be broken.
Thus, it is desirable to satisfy the conditions in Eq.~\eqref{M-conditions-B2} with the aid of right-handed quark rotation only.
To do so, we choose the correcting matrix $C_{dR}$ of the following form:
\begin{equation}
C_{dR} = \mmmatrix{1}{0}{0}{0}{c_\theta e^{i\alpha}}{s_\theta e^{i\beta}}{0}{-s_\theta e^{-i\beta}}{c_\theta e^{-i\alpha}}\,.
\end{equation}
Denoting the elements of $M_d^{(gen.)}$ as $\tilde g_{ij}$, we rewrite the conditions as
\begin{eqnarray}
&& \tilde g_{22}^* s_\theta e^{-i\beta} + \tilde g_{23}^* c_\theta e^{i\alpha} =
\tilde g_{12} s_\theta e^{i\beta} + \tilde g_{13} c_\theta e^{-i\alpha} \,,\nonumber\\
&& \tilde g_{32}^* s_\theta e^{-i\beta} + \tilde g_{33}^* c_\theta e^{i\alpha} =
\tilde g_{32} s_\theta e^{i\beta} + \tilde g_{33} c_\theta e^{-i\alpha} \,.
\label{appendix-B2-theta}
\end{eqnarray}
In each line, we group terms with $s_\theta$ and $c_\theta$, then divide the two lines, eliminating $\theta$.
After that, we regroup the resulting equation separating terms with $e^{i\beta}$ and $e^{-i\beta}$:
\begin{eqnarray}
&&\ e^{i\beta}\left[\tilde g_{12}\left(\tilde g_{33}^* e^{i\alpha} - \tilde g_{33} e^{-i\alpha}\right)
- \tilde g_{32}\left(\tilde g_{23}^* e^{i\alpha} - \tilde g_{13} e^{-i\alpha}\right) \right] \nonumber\\[1mm]
&=&
e^{-i\beta}\left[\tilde g_{22}^*\left(\tilde g_{33}^* e^{i\alpha} - \tilde g_{33} e^{-i\alpha}\right)
- \tilde g_{32}^*\left(\tilde g_{23}^* e^{i\alpha} - \tilde g_{13} e^{-i\alpha}\right) \right]\,.
\label{appendix-B2-beta}
\end{eqnarray}
The solution to this equation on $\beta$ exists, if and only if the absolute values of the two long brackets here are non-zero and equal.
This equality gives a single equation on $\alpha$, which can be represented in a compact form as
\begin{equation}
a_1 \cos(2\alpha + \psi_1) + a_2\cos(2\alpha + \psi_2) = c\,,\label{appendix-B2-eqa1}
\end{equation}
where
\begin{eqnarray}
&&a_1 = 2 |y_1| |z_1|\,, \quad \psi_1 = \arg(y_1 z_1^*)\,,\quad \mbox{with}\quad
y_1 = \tilde g_{12} \tilde g_{33}^* - \tilde g_{32} \tilde g_{23}^*\,, \quad
z_1 = \tilde g_{12} \tilde g_{33} - \tilde g_{32} \tilde g_{13}\,,\nonumber\\
&&a_2 = - 2 |y_2| |z_2|\,, \quad \psi_2 = \arg(y_2 z_2^*)\,,\quad\mbox{with}\quad
y_2 = \tilde g_{22}^* \tilde g_{33}^* - \tilde g_{32}^* \tilde g_{23}^*\,, \quad
z_2 = \tilde g_{22}^* \tilde g_{33} - \tilde g_{32}^* \tilde g_{13}\,,\nonumber\\
&& c = |y_1|^2 + |z_1|^2 -  |y_2|^2 - |z_2|^2\,.
\end{eqnarray}
This equation is further transformed to
\begin{equation}
a \cos 2\alpha  + b\sin 2\alpha  = c\,,\label{appendix-B2-eqa2}
\end{equation}
where
\begin{equation}
a = a_1 \cos\psi_1 + a_2 \cos\psi_2\,, \quad
b = -a_1 \sin\psi_1 - a_2 \sin\psi_2\,.
\end{equation}
Now, Eq.~\eqref{appendix-B2-eqa2} has solutions only if
\begin{equation}
a^2 + b^2 = a_1^2 + a_2^2 + 2a_1 a_2 \cos(\psi_1 - \psi_2) \ge c^2\,.\label{appendix-B2-eqa3}
\end{equation}
Thus, the procedure for generating a viable case $B_2$ is as follows.
We generate random $\tilde g_{ij}$, compute $a$, $b$, $c$ and check
if the inequality \eqref{appendix-B2-eqa3} is satisfied.
If not, we just interrupt evaluation and pick up another random point.
Once we find a point which satisfies condition \eqref{appendix-B2-eqa3},
we write the two solutions of Eq.~\eqref{appendix-B2-eqa2} as
\begin{equation}
\cos 2\alpha  = \frac{ac \pm b \sqrt{a^2+b^2-c^2}}{a^2+b^2}\,, \quad
\sin 2\alpha  = \frac{bc \mp a \sqrt{a^2+b^2-c^2}}{a^2+b^2}\,.\label{appendix-B2-eqa4}
\end{equation}
Once the solution for $\alpha$ is found, we insert it into Eq.~\eqref{appendix-B2-beta}
and check whether the long brackets are non-zero. If they are zero, we drop the evaluation
and pick another set of random $\tilde g_{ij}$.
Once the long brackets are non-zero, we calculate $\beta$. The last step is to insert $\alpha$ and $\beta$ into \eqref{appendix-B2-theta}
to get $\theta$. Thus, all parameters of the correcting matrix $C_{dR}$ are determined.
Once again, we arrive at a viable model with the quark rotation matrices $V_{dL} = V^{(gen.)}_{dL}$
and $V_{dR} = C_{dR}^\dagger V^{(gen.)}_{dR}$.
The same can be done in the up-quark sector without disturbing the CKM matrix.

\subsection{Case $A$}\label{appendix-caseA}

The mass matrix of case $A$ is proportional to $\Gamma_1$, which has a very constrainted structure, see Eq.~\eqref{caseA}.
It has 9 real degrees of freedom, much fewer than the other cases.
In fact, there exists a transformation $R$ which brings this matrix to the real form:
\begin{equation}
R^\dagger \Gamma_1 R = \Gamma_1^{(r.)}\,, \quad
R = \mmmatrix{1/\sqrt{2}}{i/\sqrt{2}}{0}{1/\sqrt{2}}{-i/\sqrt{2}}{0}{0}{0}{1}\,.
\end{equation}
The matrix $\Gamma_1^{(r.)}$ is a general real $3\times 3$ matrix without any further constraints.
Thus, the mass matrix of case $A$ can be produced from $D_d$ using arbitrary orthogonal matrices
in the left and right-handed quark spaces:
\begin{equation}
M_d^0 = R\cdot O_{dL} D_d O^\dagger_{dR}\cdot R^\dagger\,, \quad O_{dL}, O_{dR} \in O(3)\,.
\end{equation}
This implies $V_{dL} = R O_{dL}$ and $V_{dR} = R O_{dR}$,
solving the inversion problem for case $A$.

If we combine case $A$ in the down sector with case $B_2$ in the up sector,
then we first compute $M_d^0$, obtain $V_{dL}$, which is non-generic, then compute $V_{uL} = V_{dL}V_{\rm CKM}^\dagger$,
and follow the above two-step procedure to get the correct $B_2$ matrix.
The same procedure is applied to the combination $(B_2, A)$.
The only difficulty seems to arise for case $(A, A)$, as we are forced to use special matrices $V_{dL}$ and $V_{uL}$.
However, we immediately conclude that in this case
$V_{uL}^\dagger V_{dL} = O_{uL}^T O_{dL}$ is purely real and cannot match the experimentally known CKM matrix.
This means that case $(A, A)$ is unphysical as it misses the $CP$-violating phase of the CKM matrix.

\subsection{Case $B_3$}

In case $B_3$, the mass matrix constructed from $\Gamma_i$ given in Eqs.~\eqref{caseB3}
has a unique feature: its first $2\times 2$ block is proportional to a unitary matrix:
\begin{equation}
(M_d^0)_{2\times 2} \propto \mmatrix{g_{11}}{g_{12}}{-g_{12}^*}{g_{11}^*} = \bar g \cdot {\cal U}\,,
\quad \mbox{where}\  \bar g = \sqrt{|g_{11}|^2+|g_{12}|^2}\,, \quad {\cal U} \in SU(2)\,.
\label{appendix-caseB3-unitary}
\end{equation}
If one tries to follow the above two-step strategy and begins with a generic $M_d^{(gen.)}$,
it will be impossible to reach this form by employing block-diagonal $V_{dL}$ and $V_{dR}$.
One must use generic $SU(3)$ rotations.

Instead of trying to find the correcting matrices analytically, we resorted to the fully numerical procedure.
Namely, we use the diagonal $D_d$ and parametrize the generic unitary
$V^{(gen.)}_{dL}$ and $V^{(gen.)}_{dR}$ using the Chau-Keung form:
\begin{eqnarray}
V &=& \mmmatrix{e^{i\psi_1}}{0}{0}{0}{e^{i\psi_2}}{0}{0}{0}{e^{i\psi_3}} \times\nonumber\\
&\times &
\mmmatrix{1}{0}{0}{0}{c_{23}}{s_{23}}{0}{-s_{23}}{c_{23}}
\mmmatrix{c_{13}}{0}{s_{13}e^{-i\delta}}{0}{1}{0}{-s_{13}e^{i\delta}}{0}{c_{13}}
\mmmatrix{c_{12}}{s_{12}}{0}{-s_{12}}{c_{12}}{0}{0}{0}{1}\times\nonumber\\
&\times &
\mmmatrix{e^{i\varphi_1}}{0}{0}{0}{e^{i\varphi_2}}{0}{0}{0}{e^{i\varphi_3}}\,.\label{general-U3}
\end{eqnarray}
%Here, the three rotation angles are $\theta_{23}$, $\theta_{13}$, and $\theta_{12}$.
Since the simultaneous shift of all $\psi_i$ and all $\varphi_i$ does not change the result,
we can set one of the rephasing angles to zero, for example $\psi_3 = 0$, arriving at 9 free parameters per matrix.
%If we set $\theta_{13} = \theta_{23} = 0$, and properly arrange the phases, we recover the block diagonal form \eqref{param-block}.
%Thus, in order to explore matrices close to the block diagonal form, we need to consider small angles $theta_{13}$, $\theta_{23}$.
Treating the 18 angles as independent free parameters,
we start with a random seed point and numerically search for their values which satisfy the relations characteristic for case $B_3$.
With the multidimensional parameter space, this is done easily;
the relative accuracy achieved is $10^{-10}$ or better.

For case $(B_3,B_3)$, we perform this numerical search simultaneously in the up and down sectors,
making sure that the left-handed rotation matrices produce the CKM matrix.
Once again, all the conditions are easily satisfied, and in reasonable time we can generate
thousands of viable parameter space points for case $(B_3,B_3)$ with all quark masses and mixing
matching their experimental values.

%%%%%%%%%%%%%%%%%%%%%%%%%%%%%%%%%%%%%%%%%%%%%%%%%%%%%%%%%%%%%%%%%%%%%%%%%%%%
%%%%%%%%%%%%%%%%%%%%%%%%%%%%%%%%%%%%%%%%%%%%%%%%%%%%%%%%%%%%%%%%%%%%%%%%%%%%

\end{document}